\documentclass[aps,prd,
superscriptaddress,
showpacs,
preprintnumbers,
bibnotes,
amsmath,
amssymb,
floatfix,
]{revtex4-2}
\usepackage[utf8]{inputenc}
\usepackage{array}
\usepackage[normalem]{ulem}
\newcolumntype{L}[1]{>{\raggedright\let\newline\\\arraybackslash\hspace{0pt}}m{#1}}
\newcolumntype{C}[1]{>{\centering\let\newline\\\arraybackslash\hspace{0pt}}m{#1}}
\newcolumntype{R}[1]{>{\raggedleft\let\newline\\\arraybackslash\hspace{0pt}}m{#1}}

\DeclareMathOperator{\Cov}{Cov}

\usepackage[caption=false]{subfig}

\usepackage{graphicx}
\usepackage{dcolumn}
\usepackage{bm}
\usepackage[usenames,dvipsnames]{color}
\usepackage{verbatim}
\usepackage{amsfonts}
\usepackage{longtable}
\usepackage{natbib}
\usepackage{hyperref}
\usepackage{dcolumn}

\DeclareMathOperator{\tr}{Tr}
\newcolumntype{d}[1]{D{;}{.}{#1}}
\newcommand{\tf}{t_f}

\graphicspath{{figs/}}
\def\slashed#1{\kern+0.10em /\kern-0.50em #1}

\newcommand\bnl{Physics Department, Brookhaven National Laboratory, Upton, NY 11973, USA}
\newcommand\bnlccs{Computational Science Initiative, Brookhaven National Laboratory, Upton, NY 11973, USA}
\newcommand\cu{Physics Department, Columbia University, New York, NY 10027, USA}

\newcommand\riken{RIKEN-BNL Research Center, Brookhaven National Laboratory, Upton, NY 11973, USA}
\newcommand\regensburg{Fakult\"at f\"ur Physik, Universit\"at Regensburg, Universit\"atsstra{\ss}e 31, 93040 Regensburg, Germany}
\newcommand\edinb{School of Physics and Astronomy, The University of Edinburgh, Edinburgh EH9 3FD, UK}
\newcommand\uconn{Physics Department, University of Connecticut, Storrs, CT 06269-3046, USA}
\newcommand\soton{School of Physics and Astronomy, University of Southampton,  Southampton SO17 1BJ, UK}

\newcommand\cern{CERN, Theoretical Physics Department, Geneva, Switzerland}
\newcommand\mib{Dipartimento di Fisica, Universit\'a di Milano-Bicocca, Piazza della Scienza 3, I-20126 Milano, Italy}
\newcommand\infn{INFN, Sezione di Milano-Bicocca, Piazza della Scienza 3, I-20126 Milano, Italy}
\newcommand{\ucb}{University of California, Berkeley, CA 94720, USA}
\newcommand{\lbnl}{Lawrence Berkeley National Laboratory, Berkeley, CA 94720, USA}
\newcommand{\eic}{Electron-Ion Collider, Brookhaven National Laboratory, Upton, NY 11973, USA}
\newcommand{\cpthree}{CP$^3$-Origins \& Department of Mathematics and Computer Science, University of Southern Denmark, Campusvej 55, 5230 Odense M, Denmark}

\begin{document}
\title{An update of Euclidean windows of the hadronic vacuum polarization}
\author{T.~Blum}\affiliation{\uconn}
\author{P.~A.~Boyle}\affiliation{\bnl}\affiliation{\edinb}
\author{M.~Bruno}\affiliation{\mib}\affiliation{\infn}
\author{D.~Giusti}\affiliation{\regensburg}
\author{V.~G\"ulpers}\affiliation{\edinb}
\author{R.~C.~Hill}\affiliation{\edinb}
\author{T.~Izubuchi}\affiliation{\bnl}\affiliation{\riken}
\author{Y.-C.~Jang}\affiliation{\eic}\affiliation{\cu}
\author{L.~Jin}\affiliation{\uconn}\affiliation{\riken}
\author{C.~Jung}\affiliation{\bnl}
\author{A.~J\"uttner}\affiliation{\cern}\affiliation{\soton}
\author{C.~Kelly}\affiliation{\bnlccs}
\author{C.~Lehner}\thanks{Corresponding author}\email{christoph.lehner@ur.de}\affiliation{\regensburg}
\author{N.~Matsumoto}\affiliation{\riken}
\author{R.~D.~Mawhinney}\affiliation{\cu}
\author{A.~S.~\surname{Meyer}}\affiliation{\ucb}\affiliation{\lbnl}
\author{J.~T.~Tsang}\affiliation{\cern}\affiliation{\cpthree}

\collaboration{RBC and UKQCD Collaborations}
\noaffiliation

\date{\today}


\preprint{CERN-TH-2023-010}


\begin{abstract}
  We compute the standard Euclidean window of the hadronic vacuum polarization using multiple independent blinded analyses.  We improve the continuum and infinite-volume extrapolations of the dominant quark-connected light-quark isospin-symmetric contribution and address additional sub-leading systematic effects from sea-charm quarks and residual chiral-symmetry breaking from first principles.  We find $a_\mu^{\rm W} = 235.56(65)(50) \times 10^{-10}$, which is in $3.8\sigma$ tension with the recently published dispersive result of $a_\mu^{\rm W} = 229.4(1.4) \times 10^{-10}$ \cite{Colangelo:2022vok} and in agreement with other recent lattice determinations.  We also provide a result for the standard short-distance window.  The results reported here are unchanged compared to our presentation at the Edinburgh workshop of the g-2 Theory Initiative in 2022 \cite{EdinburghTalk}.
\end{abstract}

\maketitle


\section{Introduction}
The anomalous magnetic moment of the muon $a_\mu$ is defined as the
relative deviation of the muon's Land\'e factor $g_\mu$ from Dirac's relativistic
quantum mechanics result, $a_\mu = g_\mu / 2 - 1$.  It
is one of the most precisely determined quantities in particle
physics and has exhibited a persistent tension between the experimentally measured
value and the Standard Model theory result.

In order to reduce the experimental uncertainties, substantial efforts
are currently undertaken at Fermilab (E989) and planned at J-PARC (E34) \cite{Abe:2019thb}.
In 2021 the Fermilab experiment released first results \cite{Muong-2:2021ojo}
confirming the previously best result obtained by the BNL E821 experiment
\cite{Bennett:2006fi} and reducing the experimental
uncertainty from 0.54~ppm to 0.46~ppm.  Over the next few years, the Fermilab experiment aims to reduce the uncertainty further to approximately 0.14~ppm \cite{Carey:2009zzb}.

The Standard Model result provided by the Muon g-2 Theory Initiative \cite{Aoyama:2020ynm,Aoyama:2012wk,Aoyama:2019ryr,Czarnecki:2002nt,Gnendiger:2013pva,Davier:2017zfy,Keshavarzi:2018mgv,Colangelo:2018mtw,Hoferichter:2019mqg,Davier:2019can,Keshavarzi:2019abf,Kurz:2014wya,Melnikov:2003xd,Masjuan:2017tvw,Colangelo:2017fiz,Hoferichter:2018kwz,Gerardin:2019vio,Bijnens:2019ghy,Colangelo:2019uex,Blum:2019ugy,Colangelo:2014qya} currently has an uncertainty of 0.37~ppm and is in $4.2\sigma$ tension with the experimental value.  A further reduction of the theory uncertainty by at least
a factor of two is therefore needed \cite{Colangelo:2022jxc} to match the expected experimental progress over the next few years.  More than $90\%$ of the theory uncertainty is due to
the leading-order hadronic vacuum polarization (HVP) contribution such that a reduction of its uncertainty is particularly pressing.

The leading-order HVP contribution $a_\mu^{\rm HVP~LO}$ can be related to $e^+e^-$ decays using a dispersion relation such that, to the degree that there is no new physics
in $e^+e^-$ decays, it can be used to represent the Standard Model theory result.  The Muon g-2 Theory Initiative result quoted above uses
this method to determine the HVP contribution.  One can also relate the HVP contribution to hadronic $\tau$ decays, however, this requires
precise first-principles knowledge of the needed isospin rotation.  Our collaboration is working on such a calculation \cite{Bruno:2018ono}
and we will report on related progress in a separate publication.  Finally, the HVP contribution can be computed from first principles
using systematically improvable lattice QCD+QED methods.

Until recently, lattice QCD+QED methods have not yet been competitive with
the precision provided by the dispersive method.  The BMW collaboration, however, has now produced a lattice QCD+QED result
with $0.8\%$ precision \cite{Borsanyi:2020mff}, which is close to the current $0.6\%$ precision of the dispersive method.  The BMW value taken by itself only leads to a $1.5\sigma$ tension for $a_\mu$.  At the same time, the BMW value for the HVP contribution is in a $2.1\sigma$
tension with the dispersive result provided by the Muon g-2 Theory Initiative.

In 2018, our collaboration introduced Euclidean window quantities \cite{RBC:2018dos}, which allow for the separation of the most
challenging short and long time-distance contributions to $a_\mu^{\rm HVP~LO}$.
The remaining standard window quantity, $a_\mu^{\rm HVP~LO~W}$, is much easier to compute at high
precision in lattice QCD+QED and can also be computed using the dispersive method \cite{RBC:2018dos,Aubin:2019usy,Borsanyi:2020mff,Colangelo:2022vok}.  The BMW collaboration's calculation of $a_\mu^{\rm HVP~LO~W}$ is in fact in $3.7\sigma$
tension with the dispersive result, which has motivated many lattice collaborations to focus on high-precision calculations of $a_\mu^{\rm HVP~LO~W}$ first in order to
clarify the situation.  In this work, we provide a significantly improved calculation of $a_\mu^{\rm HVP~LO~W}$.  We focus on the the quark-connected light-quark contribution in the isospin-symmetric limit, which accounts for almost $90\%$ of $a_\mu^{\rm HVP~LO~W}$.  Special attention is given to the continuum limit for which we replace our previous continuum
extrapolation based on a single approach using 2 lattice spacings with one based on 8 distinct approaches using 3 lattice spacings.
We perform this update using a blinding procedure with five independent analysis groups. This blinding procedure is implemented to avoid bias toward our previous computation of $a_\mu^{\rm HVP~LO~W}$ in Ref.~\cite{RBC:2018dos}, the dispersive results, or other lattice results.

This paper is organized as follows.  In Sec.~\ref{sec:methodology}, we describe our methodology before giving computational details in Sec.~\ref{sec:computationaldetails}.
In Sec.~\ref{sec:relativeblind}, we discuss blinded results and explain convergence to the final prescription to determine $a_\mu^{\rm HVP~LO~W}$.  Finally,
in Sec.~\ref{sec:unblind}, we present unblinded results and compare them to other groups' results, including data-driven ones, before concluding in Sec.~\ref{sec:conclude}.

\section{Methodology}\label{sec:methodology}
We first define the time-momentum representation in Sec.~\ref{sec:tmr}, which provides
the basis for the definition of the Euclidean windows in Sec.~\ref{sec:window}.   In Sec.~\ref{sec:iso} we define the isospin-symmetric world around which we expand.
Special care is taken such that the isospin-symmetric contribution can be compared directly with other lattice results.  In Sec.~\ref{sec:blind}, we describe our blinding procedure.

\subsection{Time-momentum representation}\label{sec:tmr}
\begin{figure}
\includegraphics[width=12cm]{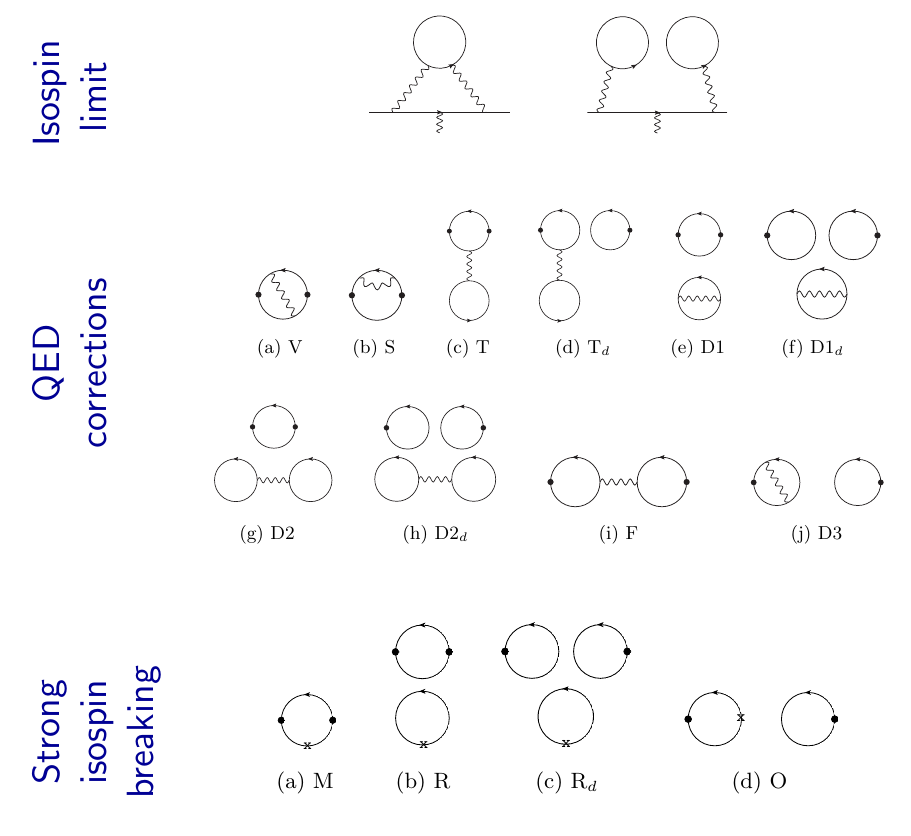}
\caption{\label{fig:diagrams} The diagrams of a complete calculation of $a_\mu^{\rm HVP~LO}$ when formulated as an expansion around an isospin-symmetric limit.
  In the isospin-symmetric limit, there is a quark-connected (left) and quark-disconnected contribution (right).  For the QED- and strong-isospin-breaking (SIB) corrections,
  we indicate the photon vertices that connect to the muon with filled dots and only show the respective sub-diagrams.  For the QED corrections, one has to enforce the exchange of gluons
  between the quark loops in diagram F to avoid double-counting of higher-order HVP contributions.  For the SIB corrections, the crosses denote scalar operator insertions to allow for
  a linear correction in the respective quark masses.
}
\end{figure}

Starting from the vector current 
$J_\mu(x) = i\sum_f Q_f \overline{\Psi}_f(x) \gamma_\mu \Psi_f(x)$ with fractional electric charge $Q_f$ and sum over quark flavors $f$
we may write
\begin{align}
  a^{\rm HVP~LO}_\mu &= \sum_{t=0}^\infty w_t C(t)
  \label{eq:tmr}
\end{align}
with correlator
\begin{align}
  C(t) = \frac13 \sum_{\vec{x}}\sum_{j=0,1,2} \langle J_j(\vec{x},t) J_j(0) \rangle \,,
\end{align}
where the weights $w_t$ capture the photon and muon part of the HVP diagrams.  A complete list of diagrams is given in Fig.~\ref{fig:diagrams}.
The weights can be expressed as a one-dimensional integral \cite{Bernecker:2011gh}
\begin{align}
    w_t &= 8\alpha^2\int_0^\infty dQ^2\left(\frac{\cos{(Qt)}-1}{Q^2}+\frac{1}{2}\, t^2\right)f(Q)
\label{eq:wt}
\end{align}
with
\begin{align}    
f(Q)&=\frac{m_\mu^2 Q^2 Z^3(Q)(1-Q^2 Z(Q))}{1+m_\mu^2 Q^2 Z^2(Q)}\,, &
Z(Q)&=\frac{\sqrt{Q^4+4Q^2 m_\mu^2}-Q^2}{2m_\mu^2 Q^2} \,,
\end{align} 
where $m_\mu$ is the muon mass.
Note that we sum only over non-negative $t$ in Eq.~\eqref{eq:tmr}, yielding an additional symmetry factor of two in $w_t$.
Using a lattice discretization for the photon momenta, an alternative weight
\begin{align}\label{eq:wthat}
    \hat{w}_t&=8\alpha^2\int_0^\infty dQ^2\left(\frac{\cos{(Qt)}-1}{(2\sin{Q/2})^2}+\frac{1}{2}\, t^2\right)f(Q)
\end{align}
can be defined, which gives the same value of $a^{\rm HVP~LO}_\mu$ in the continuum limit.
We use both versions to scrutinize the continuum extrapolation.

The correlator $C(t)$ is
computed in lattice QCD+QED at physical pion mass with
non-degenerate up- and down-
quark masses including up-, down-, strange-,
and charm-quark contributions.  The missing bottom-quark
contributions are estimated using perturbative QCD.

\subsection{Euclidean windows}\label{sec:window}
In the following, we suppress the leading-order HVP~LO label for brevity.
Following \cite{RBC:2018dos}, we define Euclidean windows that partition the contributions of time-slices $t$
in Eq.~\eqref{eq:tmr} into short-distance (SD), window (W), and long-distance (LD) contributions.  To make
the quantities well-defined at non-zero lattice spacing, we introduce smearing kernels with width $\Delta$.
We write
\begin{align}
  a_\mu &= a_\mu^{\rm SD} + a_\mu^{\rm W} + a_\mu^{\rm LD}\,, 
\end{align}
where
\begin{align}
  a_\mu^{\rm SD}(t_0,\Delta) &= \sum_{t=0}^\infty C(t) w_t [1 - \Theta(t,t_0,\Delta)] \,, \\ 
  a_\mu^{\rm W}(t_0,t_1,\Delta) &=  \sum_{t=0}^\infty C(t) w_t [ \Theta(t,t_0,\Delta) - \Theta(t,t_1,\Delta) ] \,, \\
  a_\mu^{\rm LD}(t_1,\Delta) &=  \sum_{t=0}^\infty C(t) w_t \Theta(t,t_1,\Delta) \,, \\
  \Theta(t,t',\Delta) &= \left[1 + \tanh\left[ (t-t')
      / \Delta \right]\right]/2 \,.
\end{align}
All contributions are well-defined individually and can be computed using lattice methods as well as dispersive methods by
relating the correlator
\begin{align}
  C(t) =\frac{1}{12 \pi^2} \int_0^\infty d(\sqrt{s}) R(s) s e^{-\sqrt{s} t}
\end{align}
to the R-ratio
\begin{align}
  R(s) =\frac{3 s}{4 \pi \alpha^2} \sigma(s,e^+e^- \to {\rm had}).
\end{align}
Within a lattice calculation, discretization effects are most severe for the SD contribution, while statistical noise and finite-volume effects are most
pronounced in the LD contribution.  The window quantity $a_\mu^{\rm W}$ has small statistical and systematic errors.

As recently argued in Ref.~\cite{Colangelo:2022vok}, the systematic study of window quantities $a_\mu^{\rm W}(t_0,t_1,\Delta)$ as a function of $t_0$ and $t_1$ is useful to constrain
energy regions within the R-ratio contributing to a possible tension between lattice and dispersive results.  First lattice results with a high resolution
in $t_0$ and $t_1$ are already available \cite{Lehner:2020crt}.
Windows with larger values of $t_0$ and $t_1$ are more sensitive to low-energy states and are useful for checking effective field theory as argued in Ref.~\cite{Aubin:2022hgm}.
A systematic study of the short-distance window $a_\mu^{\rm SD}(t_0,\Delta)$ as a function of $t_0$ is also useful as argued in Ref.~\cite{FermilabLattice:2022izv}, where the $a_\mu^{\rm SD}(t_0,\Delta)$ defined as above are called one-sided windows since $1 - \Theta(t,t_0,\Delta)=\left[1 - \tanh\left[ (t-t_0)
      / \Delta \right]\right]/2 = \Theta(t_0,t,\Delta)$.  In the current work, we focus on the short-distance and window contributions for
the standard values of $t_0=0.4$ fm, $t_1=1.0$ fm, and $\Delta=0.15$ fm \cite{RBC:2018dos}.

\subsection{Isospin-symmetric world}\label{sec:iso}
It is convenient to perform the calculation as an expansion around an isospin-symmetric point \cite{deDivitiis:2013xla,Boyle:2017gzv,RBC:2018dos,Giusti:2019xct}.  We therefore compute the diagrams of Fig.~\ref{fig:diagrams} individually.  The exact choice of the
expansion point is inconsequential for the total $a_\mu$, however, care is needed if
one attempts to compare isospin-symmetric results provided by different groups
\footnote{For a discussion of scheme ambiguities in light-meson leptonic decays, see Refs.~\cite{DiCarlo:2019thl,Boyle:2022lsi}.}.

In this work, we present results for two choices of the 
isospin-symmetric world.  The
first choice is the RBC/UKQCD18 world defined by
\begin{align}\label{eqn:rbcworld}
m_\pi&=0.135~\text{GeV}\,, &
m_K&=0.4957~\text{GeV}\,, &
m_\Omega &= 1.67225~\text{GeV} \,,
\end{align}
consistent with our previous work \cite{RBC:2018dos}.  In this
update, we also consider the effects from dynamical sea-charm quarks from first principles and therefore extend this choice by
\begin{align}\label{eqn:mds}
  m_{D_s} &=1.96847~\text{GeV} \,.
\end{align}
Since one of the main goals of this work is to scrutinize the result of Ref.~\cite{Borsanyi:2020mff}, we also consider a second choice
\begin{align}\label{eqn:bmwworld}
  m_\pi&=0.13497~\text{GeV}\,, &
  m_{ss*}&=0.6898~\text{GeV}\,, &
  w_0 &=0.17236~\text{fm} \,,
\end{align}
which we label as the BMW20 world.  The quantity $m_{ss*}$ is obtained from the ground-state energy of the quark-connected pseudoscalar $\bar{s}s$ meson two-point function.  This choice is consistent with the isospin-symmetric world defined in Ref.~\cite{Borsanyi:2020mff}.
For the sea-charm study, we adopt Eq.~\eqref{eqn:mds} also in this case.

We define these parameters to the exact values given above without additional uncertainty.
This avoids an unnecessary inflation of uncertainties when comparing isospin-symmetric lattice results.  The experimental uncertainties of the physical hadron spectrum are then taken into account when applying the isospin-breaking corrections.

To support the careful tuning of the isospin-symmetric world, we 
generated additional near-physical-pion-mass ensembles allowing for the explicit calculation of light and strange quark-mass derivatives. Our choice of discretisation and simulation parameters is summarised in Tab.~\ref{tab:ex}. We also generated ensembles with dynamical charm quarks and ensembles with varying extent of the fifth dimension of
our domain-wall fermions, $L_s$, to 
control for residual chiral-symmetry-breaking effects.
Finally, we include results at physical pion mass and a finer lattice spacing of $a^{-1} \approx 2.7$ GeV.

We determined the ensemble parameters in two ways.  First, we used the
new ensembles to obtain the quark-mass dependence of the quantities defined
in Eqs.~\eqref{eqn:rbcworld} and \eqref{eqn:bmwworld}.
We then tune the dimensionless $m_\pi/m_\Omega$ and $m_K/m_\Omega$ for the RBC/UKQCD18 world and $w_0 m_\pi$ and $w_0 m_{ss*}$ for the BMW20 world to the values provided in Eqs.~\eqref{eqn:rbcworld} and \eqref{eqn:bmwworld}.
Any of the three dimensionful values can then equivalently be used to determine the lattice spacing $a$ for a given ensemble.  For the $N_f=2+1+1$ ensembles, we also tune $m_D{_s}/m_\Omega$ for the RBC/UKQCD18 world and $w_0 m_{D_s}$ for the BMW20 world to the value provided in Eq.~\eqref{eqn:mds}.  We provide the results for the RBC/UKQCD18 world in Tab.~\ref{tab:ex}.  In addition, we also performed an update of our global fit \cite{RBC:2014ntl} for which we found consistent results.  A detailed discussion of the updated global fit will be published separately.  The two determinations of ensemble parameters were performed by disjoint sub-groups of authors.

\begin{table}
\begin{ruledtabular}
  \begin{tabular}{l|lllllllll}
  ID & $a^{-1}$/GeV & $N_f$ & $L^3 \times T \times L_s/a^4$ & $b+c$ & $am_{\rm res} \times 10^4$ & $m_\pi$/MeV & $m_K$/MeV & $m_{D_s}$/GeV & $m_\pi L$  \\\hline
  48I & $1.7312(28)$ & 2+1 & $48^3 \times 96 \times 24$ & 2 & $6.1$ & $139.32(30)$ & $499.44(88)$ & -- & 3.9 \\
  64I & $2.3549(49)$ & 2+1 & $64^3 \times 128 \times 12$ & 2 & $3.1$ & $138.98(43)$ & $507.5(1.5)$ & -- & 3.8 \\
  96I & $2.6920(67)$ & 2+1 & $96^3 \times 192 \times 12$ & 2 & $2.3$ & $131.29(66)$ & $484.5(2.3)$ & -- & 4.7 \\\hline
  1   & $1.7310(35)$ & 2+1 & $32^3 \times 64 \times 24$ & 2 & $6.3$ & $208.1(1.1)$ & $514.0(1.8)$ & -- & 3.8  \\
  2   & $1.7257(74)$ & 2+1 & $24^3 \times 48 \times 32$ & 2 & $4.6$ & $285.4(2.9)$ & $537.8(4.6)$ & -- & 4.0  \\
  3   & $1.7306(46)$ & 2+1 & $32^3 \times 64 \times 24$ & 2 & $6.5$ & $211.3(2.3)$ & $603.8(6.1)$ & -- & 3.9  \\
  4   & $1.7400(73)$ & 2+1 & $24^3 \times 48 \times 24$ & 2 & $6.2$ & $274.8(2.5)$ & $530.1(3.1)$ & -- & 3.8  \\
  5   & $1.7498(73)$ & 2+1+1 & $24^3 \times 48 \times 24$ & 2 & $6.7$ & $279.8(3.5)$ & $539.1(5.3)$ & $1.9902(69)$ & 3.8 \\
  7   & $1.7566(81)$ & 2+1+1 & $24^3 \times 48 \times 24$ & 2 & $7.9$ & $272.5(5.9)$ & $523(10)$ & $1.3882(57)$ & 3.7 \\
  A   & $1.7556(83)$ & 2+1 & $24^3 \times 48 \times 8$ & 2 & $42$ & $307.4(3.5)$ & $557.3(5.7)$ & -- & 4.2 \\\hline
  24ID & $1.0230(20)$ & 2+1 & $24^3 \times 64 \times 24$ & 4 & $23$ & $142.96(30)$ & $515.7(1.0)$ & -- & 3.4 \\
  32ID & $1.0230(20)$ & 2+1 & $32^3 \times 64 \times 24$ & 4 & $23$ & $142.96(30)$ & $515.7(1.0)$ & -- & 4.5 \\
  \end{tabular}
\end{ruledtabular}
\caption{\label{tab:ex} List of ensembles with parameters determined in the RBC/UKQCD18 isospin symmetric world.  Unless specified otherwise, the ensembles have
  Iwasaki gauge action and M\"obius \cite{Brower:2012vk} domain-wall \cite{Shamir:1993zy,Furman:1994ky} fermion sea quarks with $b-c=1$.  The parameters $b$ and $c$ are defined in Ref.~\cite{RBC:2014ntl}.  For the $N_f=2+1+1$ ensembles, the charm quarks couple to three-times $\rho=0.1$ stout smeared gauge fields
  as in Refs.~\cite{Cho:2015ffa,Boyle:2018knm}.  The scripts generating the new ensembles
are publicly available \cite{GPTensembles}.  The 24ID and 32ID ensembles have an additional DSDR term \cite{RBC:2014ntl} in the gauge action.  The 24ID and 32ID ensemble parameters are taken from Ref.~\cite{JiqunThesis}.}
\end{table}

\subsection{Blinding procedure}\label{sec:blind}
Since we provide an update of a previous result \cite{RBC:2018dos} compared to which a lower value would mean agreement with the dispersive method and a higher value would mean agreement with the lattice result of Ref.~\cite{Borsanyi:2020mff}, two values that are in $3.7\sigma$ tension with each other, we believe it is crucial to perform this update in a blinded manner.

We implement the blinding by creating modified correlators $C_b(t)$ from the unaltered correlators $C_0(t)$.  For each lattice ensemble, we use
\begin{align}\label{eqn:blind}
C_b(t) = (b_0 + b_1 a^2 + b_2 a^4) C_0(t)
\end{align}
with respective lattice spacing $a$ and random coefficients $b_0$, $b_1$, and $b_2$ that are common for each ensemble but different for each analysis group.  The parameter $b_0$ is drawn from a Gaussian distribution with mean $\mu=1.0$ and standard deviation $\sigma=0.2$.  The dimensionful parameters $b_1$ and $b_2$
are drawn from a flat distribution with maximum values of $\vert b_1 a^2 \vert = 0.05$ and $\vert b_2 a^4 \vert = 0.0025$ for our coarsest lattice cutoff $a^{-1}=1.73$ GeV.  This procedure based on three random numbers per analysis group prevents the possibility of complete unblinding based on previously shared data on the coarser two ensembles \cite{RBC:2018dos}.  The blinding factors were generated and directly applied to $C_0$ by author CL.  This process took a given seed for the random number generator as input such that only this seed and not the blinding factors were directly accessible to CL.

For the current update, we established five analysis groups (called A--E in the following), composed of non-overlapping sub-groups of authors.
 The different analysis groups were provided with the ensemble parameters 
and the respectively blinded correlator data.  They then separately decided on their
respective analysis procedures without interacting with other groups. 
The chosen methods are described in Sec.~\ref{sec:distinctmethods}.
After the groups
completed their analyses, we started a relative unblinding procedure during which two groups would jointly
discuss and scrutinize their approaches.  In this process
some important findings emerged, as described in Sec.~\ref{sec:findings}.  Based on these discussions, the collaboration
then converged on a preferred prescription that is described in Sec.~\ref{sec:preferred}.  At this point the prescription was frozen and a complete
unblinding performed.  The results are discussed in Sec.~\ref{sec:unblind}.

\section{Computational details}\label{sec:computationaldetails}
In the following, we describe in detail the computational methods used in this work.  We explain aspects of data generation as well as crucial  components of the various $a_\mu$ analyses.

\subsection{Overview of improvements}\label{sec:overviewimpro}
Compared to our previous calculation of Ref.~\cite{RBC:2018dos}, we
have made several substantial improvements.  With regard to the statistical uncertainty, we increased the statistical sample size for the correlators on ensembles 48I and 64I by a factor of four.  Improvements reducing systematic uncertainties are described in the following.

To improve the continuum extrapolation, we add a finer lattice spacing at physical pion mass with $a^{-1}=2.7$ GeV.
We also consider an additional discretization for the vector current by studying both local-conserved as well as local-local correlators.  This can be done in a cost-efficient manner as described in Sec.~\ref{sec:conserved}.  In addition, we use two different
renormalization procedures for the local vector current.  The first procedure, which we label $Z_V$, follows Ref.~\cite{RBC:2014ntl} and uses that the expectation value of the charge operator in a pion state equals one.  The second procedure, which we label $Z_V^\star$, uses the ratio of local-conserved to local-local correlators interpolated to fixed Euclidean time $t^\star$ to define the current normalization.  The particular choice of $t^\star$ is described in Sec.~\ref{sec:distinctmethods}.  Finally, we use two
different weight functions $w_t$ and $\hat{w}_t$, see Eqs.~\eqref{eq:wt} and \eqref{eq:wthat}, at a given lattice spacing.  This gives a total of $3 \times 2 \times 2 \times 2 = 24$ data points to study the continuum extrapolation, which  improves our previous extrapolation based on two data points.

To reduce parametric uncertainties, we generated new near-physical pion- and kaon-mass ensembles to calculate parametric derivatives with respect to quark masses.  In Sec.~\ref{sec:mfder}, we also show how to obtain parametric derivatives inspired by master-field methodology \cite{Luscher:2017cjh}.

We previously estimated the missing sea-charm effects using perturbative QCD \cite{RBC:2014ntl}.  For this update, we have generated new ensembles with dynamical
charm quarks, which we match to our $N_f=2+1$ ensembles as described in Sec.~\ref{sec:seacharm}.

Domain-wall fermions exhibit only small chiral symmetry breaking which is commonly quantified using the residual mass $m_{\rm res}$ \cite{Furman:1994ky,Brower:2004xi}.  For this reason, a very small linear discretization error is allowed.  We previously neglected
such effects but have now generated new ensembles with different extents of the fifth dimension $L_s$
to quantify them from first principles.

Since we only have a small number of configurations for the new 96I ensemble, we also
investigate a new five-dimensional master-field statistical error estimate in Sec.~\ref{sec:mferr} to considerably reduce the uncertainty on our estimate of statistical variance concerning this ensemble.

\subsection{Local- and conserved-current correlators}\label{sec:conserved}
In addition to the local lattice vector current $J_\mu$, which we denote in the following as $J^{\rm l}_\mu$, we consider the conserved lattice vector current $J^{\rm c}_\mu$ as defined in Ref.~\cite{RBC:2014ntl}.  We consider the  correlators
\begin{align}
  C^{ab}(t) = \frac13 \sum_{\vec{x}}\sum_{j=0,1,2} \langle J^b_j(\vec{x},t) J^a_j(0) \rangle 
\end{align}
in the local-local ($C^{\rm ll}$) and local-conserved ($C^{\rm lc}$) versions.
After performing the fermionic Wick contraction, the source is always local and the sink varies between local and conserved.  The contraction code is publicly available \cite{GPThvp}.  It uses an all-mode-averaging procedure \cite{DeGrand:2004wh,Bali:2009hu,Blum:2012uh,Shintani:2014vja} combined with additional averaging of the low-low component of the correlator \cite{RBC:2018dos}.  Our approach again relies on approximating the low-mode space on a coarse grid as introduced in Ref.~\cite{Clark:2017wom}.  For the 96I ensemble, this yields a reduction of data volume by a factor of 30.  This is crucial not just for data storage but also for the computational performance of low-mode estimates due to the reduced memory-transfer requirements.

\begin{figure}
\includegraphics[width=10cm]{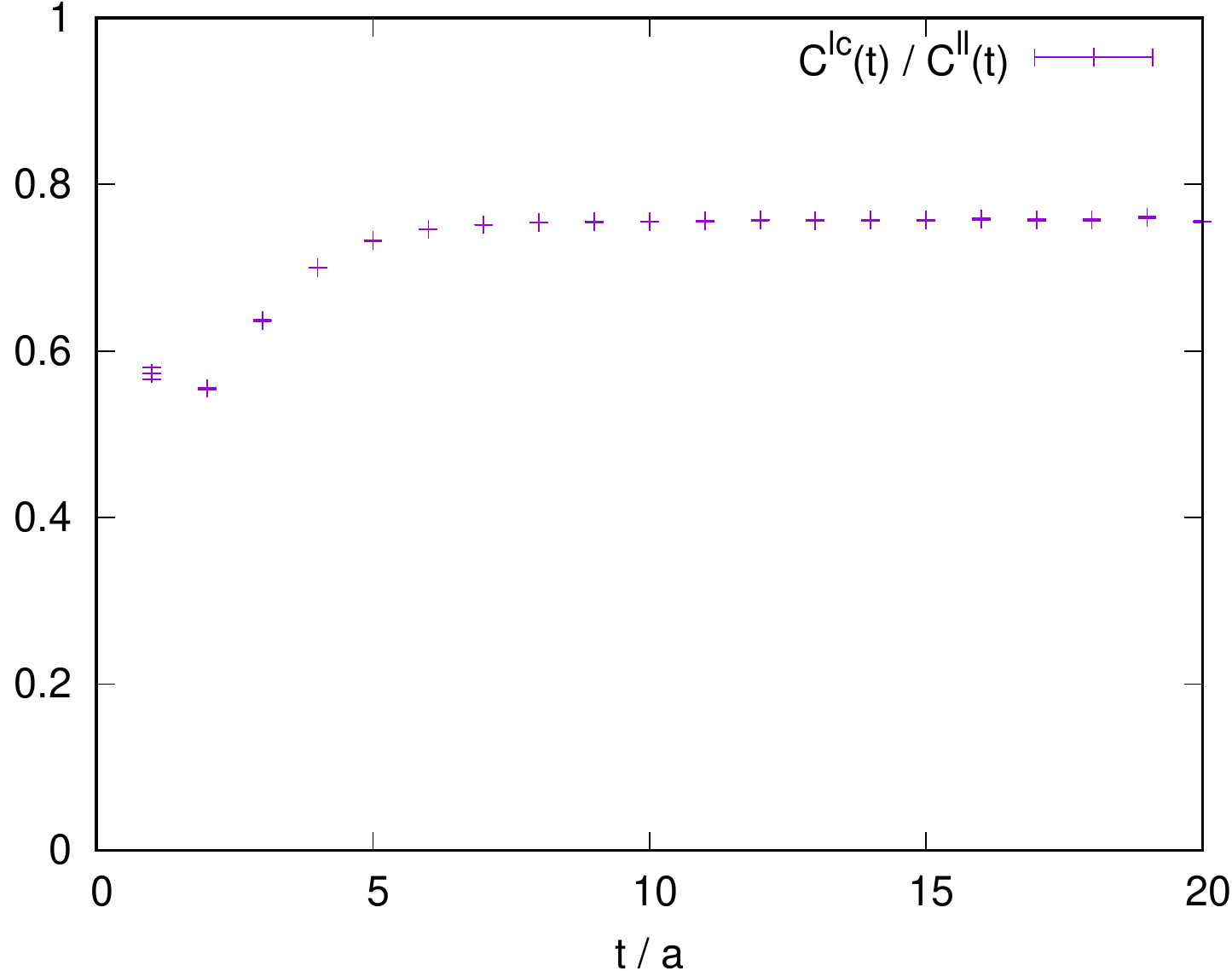}
\caption{\label{fig:ratiolcll}Ratio $C^{\rm lc}(t) / C^{\rm ll}(t)$ as a function of Euclidean time $t$ on the 96I ensemble.}\end{figure}

For a given point source, the local-local and local-conserved correlators are highly correlated.  We therefore compute the ratio $C^{\rm lc} / C^{\rm ll}$ using only a few correlated source positions and multiply this ratio with our full-statistics estimator of $C^{\rm ll}$ to obtain our estimator for $C^{\rm lc}$.  In Fig.~\ref{fig:ratiolcll}, we plot
the ratio for the 96I ensemble.

For the 96I ensemble an additional improvement was made.  For this ensemble, we generate a data set in which two source positions at time-slice $t$ and $t + 96$ are combined with a $Z_2$ number.  For short and intermediate distances, this effectively doubles our statistics at the same cost.  A second lower-statistics single time-slice data set is provided to account for the effects of the backwards propagation of the additional time slice.

Finally, all correlators are provided with identical valence- and sea-quark masses.  In this manner, we can perform a purely unitary data analysis.  For the 64I ensemble, however, for historical reasons the eigenvectors were generated for a partially-quenched mass $am=0.0006203$ instead of the unitary mass $am=0.0006780$ \cite{RBC:2014ntl}.  For this reason, a small additional correlated data set was generated at $am=0.001774$
such that the unitary correlators can be obtained by
\begin{align}\notag
      a_\mu(0.0006780) &= a_\mu(0.0006203) + (0.0006780-0.0006203) \frac{a_\mu(0.001774)-a_\mu(0.0006203)}{0.001774-0.0006203} \\
      &\approx a_\mu(0.0006203) + \frac{a_\mu(0.001774)-a_\mu(0.0006203)}{20} \,.
\end{align}
Non-linear effects in the small quark-mass shift are negligible for the precision goals of the present calculation.

\subsection{Sea-charm effects}\label{sec:seacharm}
In this work, we estimate the effects of sea-charm quarks from first principles.  Most of our ensembles have $N_f=2+1$ sea quarks with an isospin-symmetric up- and down-quark pair and an additional strange quark.
To study the sea-charm effects from first principles, we have generated additional $N_f=2+1+1$ ensembles with different charm masses to separate the physical effects from a modification of discretization errors.  We list 
the ensemble parameters in Tab.~\ref{tab:ex}.

\begin{figure}
\includegraphics[width=10cm]{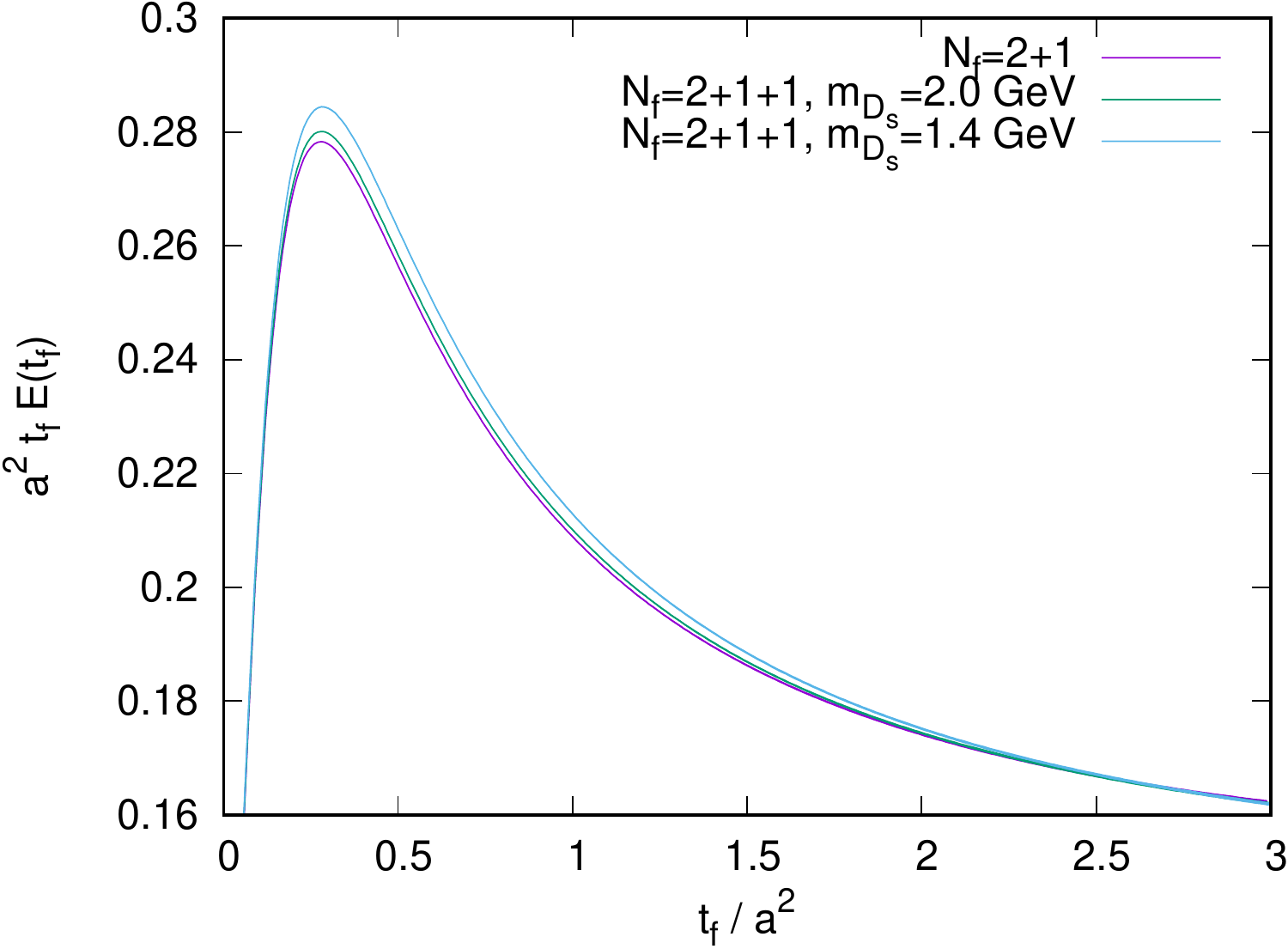}
\caption{\label{fig:charmwfmatch} Wilson-flowed energy density $E(\tf)$ multiplied with the flow-time $\tf$ for $N_f=2+1$ and $N_f=2+1+1$ ensembles.  The small statistical uncertainties for each line are shown as an error band.}
\end{figure}

We match the $N_f=2+1$ and $N_f=2+1+1$ ensembles to the same pion and kaon masses and the Wilson-flowed~\cite{Luscher:2010iy} energy density at long-distance.   In Fig.~\ref{fig:charmwfmatch}, we show $\tf E(\tf)$ with flow-time $\tf$ and Wilson-flowed energy density $E(\tf)$ for the nominal ensemble 4, 5, and 7 of Tab.~\ref{tab:ex}.  At shorter distances, we observe a clear signal of charm effects in the energy density.  For the lighter charm mass, this effect extends to longer distances.  We plot $\tf E(\tf)$ instead of the dimensionless $\tf^2 E(\tf)$ since all plotted ensembles share the same lattice spacing and the interesting features are better highlighted in this way.

We use these matched ensembles to measure the sea-charm contributions to the HVP.  We do this in particular for the short-distance window, where most of the effect should appear.  The exact approach used by the different analysis groups is explained in Sec.~\ref{sec:distinctmethods}.

\subsection{Five-dimensional master-field statistical errors}\label{sec:mferr}
For the 96I ensemble, we currently only have 33 gauge field configurations in contrast to the 64I and 48I ensembles for which we have 238 and 386 gauge field configurations, respectively.  In order to obtain a reliable statistical-error estimate on the 96I ensemble, we have performed a slightly modified master-field error analysis \cite{Luscher:2017cjh}.  In our approach, we improve the covariance estimate by considering a five-dimensional master field with Markov time as an additional fifth dimension.  We expect exponential locality in the fifth dimension governed by the eigen-modes of the Markov transition matrix and in the four space-time dimensions governed by the eigen-modes of the QCD Hamiltonian.

For an observable $O_{\tau,x}$ with Markov time $\tau$ and space-time coordinate $x$, we consider the
statistical average
\begin{align}
 O = \frac{1}{\vert \mathrm{V} \vert} \sum_{(\tau,x) \in \mathrm{V}} O_{\tau,x}
\end{align}
with set V that contains all tuples $(\tau,x)$ for which the observable was determined.
Note that we explicitly allow for sparse sampling in space-time as well as Markov time.  The covariance of two such observables $O$ and $O^\prime$ is then given by
\begin{align}\label{eqn:defcov}
  \Cov_{\tau_c,x_c}(O,O^\prime)  &\equiv  \frac1{\vert \mathrm{V} \vert \vert \mathrm{V}^\prime \vert} \sum_{
  \stackrel{(\tau,x) \in \mathrm{V},(\tau^\prime,x^\prime) \in \mathrm{V}^\prime,}{\vert x - x^\prime \vert \leq x_c,\vert \tau - \tau^\prime \vert \leq \tau_c}} \bigl(\langle O_{\tau,x} O^\prime_{\tau^\prime,x^\prime} \rangle - \langle O_{\tau,x} \rangle \langle O^\prime_{\tau^\prime,x^\prime} \rangle \bigr)
\end{align}
and studying $\Cov_{\tau_c,x_c}(O,O^\prime)$ as a function of $\tau_c$ and $x_c$ to identify a plateau for large $\tau_c$ and $x_c$.  In practice, we estimate $\Cov_{\tau_c,x_c}(O,O^\prime)$ based on a given set of gauge configurations, which adds an error suppressed by the inverse square root of the number of sampled five-dimensional points.  In comparison, the Jackknife estimator has an uncertainty suppressed by the inverse square root of the number of gauge configurations, such that its uncertainty is generally much larger. 
The distance $\vert x - x^\prime \vert$ takes the field boundary conditions into account, i.e., for periodic boundary conditions, we consider the shortest distance between mirror images.

\begin{figure}
\includegraphics[width=10cm]{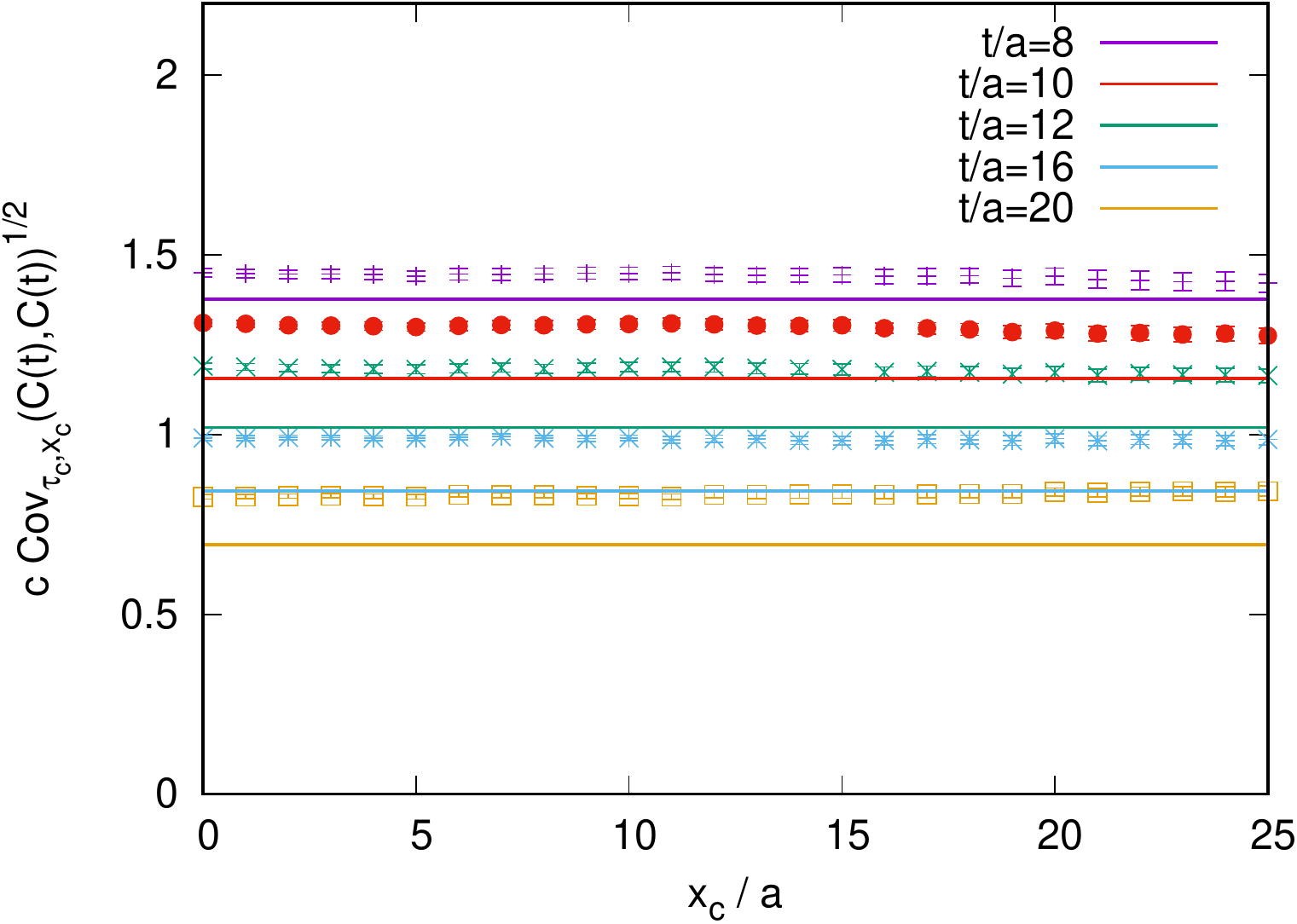}
\caption{\label{fig:me96Ierr} The statistical uncertainty of $C(t)$ determined by Eq.~\eqref{eqn:defcov}
 multiplied with a blinding factor $c$ determined by the five-dimensional master-field approach (individual data points) compared to the Jacknife estimate (solid lines).  For these estimates, we use randomly selected 660 point sources per 33 configurations on the 96I ensemble.  Due to the sparseness of our measurement setup, we observe a plateau in $x_c$ starting essentially from the smallest value.  The plot is made after having established a plateau in $\tau_c$.}
\end{figure}

For arbitrarily sparse V, the various $O_{\tau,x}$ are effectively all statistically independent such that we expect a plateau already for very small $\tau_c$ and $x_c$.  In general, just before reaching the gauge noise limit, the plateaus still start early in $x_c$.  Conversely, a rising behavior in $x_c$ signals that our sample points are significantly correlated.  We tune the sampling of our vector correlators to be such that we almost reach the gauge noise limit and therefore plateaus are reached for modest values of $x_c$.  In Fig.~\ref{fig:me96Ierr}, we compare the statistical uncertainty of $C(t)$ on the 96I ensemble determined by the five-dimensional master-field approach to the Jackknife estimate.

\subsection{Master-field parametric derivatives}\label{sec:mfder}
In order to tune the $N_f=2+1+1$ ensembles described in Sec.~\ref{sec:seacharm}, we found the
master-field formalism useful to get initial estimates of parametric derivatives with respect to
the gauge-action parameter $\beta$ as well as the sea-charm mass.  To simplify the discussion, we set $a=1$ in this sub-section.

Consider a general gauge action
\begin{align}\label{eq:action}
 S = - \beta \frac{N_d(N_d-1)}{2} \sum_x A_x
\end{align}
with space-time dimension $N_d$ and field of Wilson loops $A_x$ anchored at a point $x$.
It is not crucial how we exactly identify the location $x$ as long as the coordinate behaves properly under translations of the system.  One can then show that for a general observable $O$ in $N_d=4$ without explicit $\beta$ dependence,
\begin{align}
\partial_{\beta} \langle O \rangle = 6 \lim_{x_c \to \infty} \Cov_{0,x_c}(O,A)\,,
\end{align}
with $\Cov_{0,x_c}$ defined in Eq.~\eqref{eqn:defcov}.  Setting $O$ to the Wilson-flowed energy density $E(\tf)$, e.g., allows us to determine the $\beta$-derivative of the Wilson-flow scales $t_0$ and $w_0$~\cite{Luscher:2010iy,Borsanyi:2012zs}.

\begin{figure}
\includegraphics[width=7.5cm,page=1]{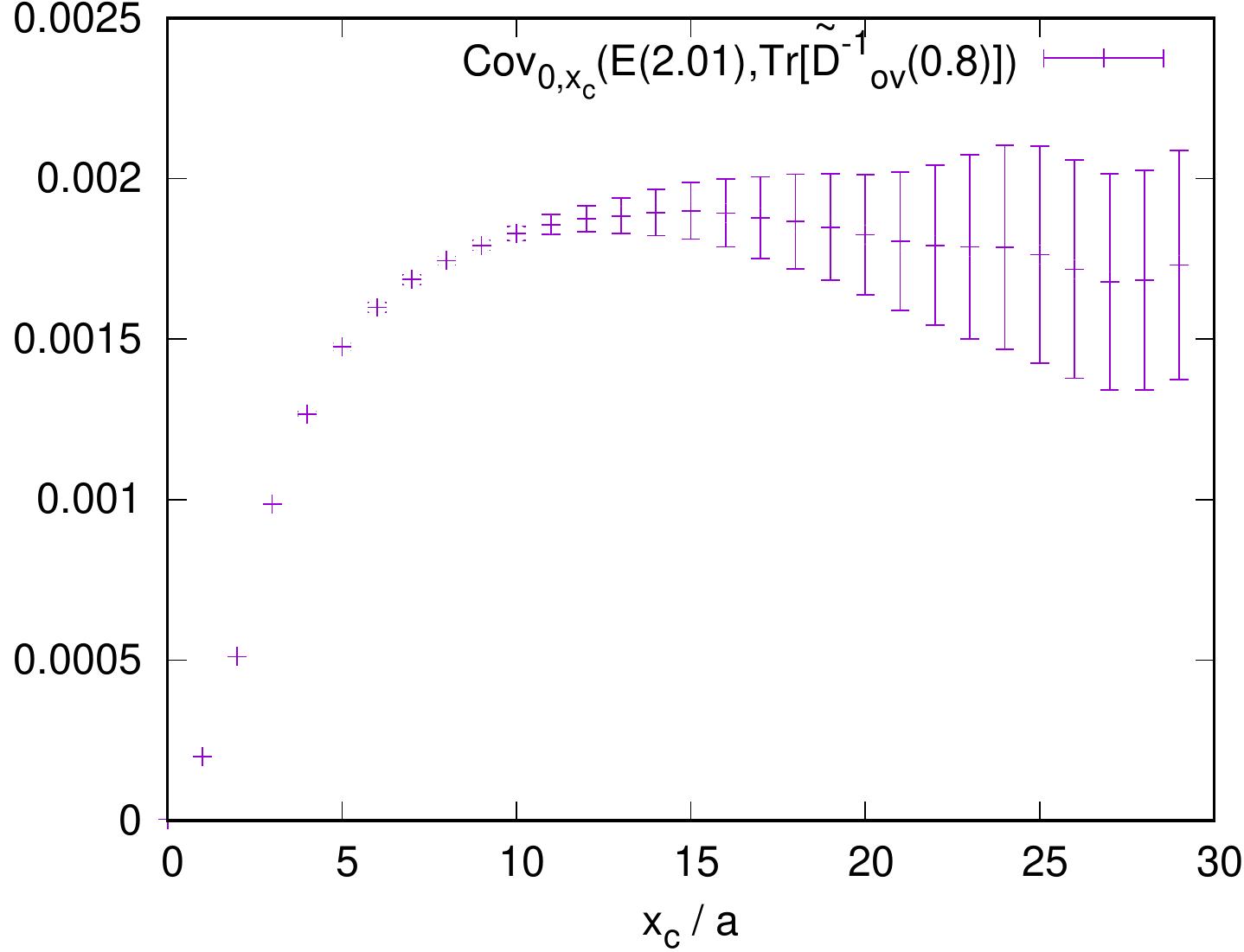}\hspace{1cm}\includegraphics[width=7.5cm,page=2]{figs/mfplots-crop}
\caption{\label{fig:mfpar} We plot for the 96I ensemble
$\Cov_{0,x_c} (E(t_f), \tr[ \tilde{D}^{-1}_{\rm ov}(0.8)])$ on the left and
$\Cov_{0,x_c} (E(t_f), A)$ on the right for $t_f=2.01\approx t_0/2$.  The Wilson-loop field $A$ is defined in Eq.~\eqref{eq:action}.}
\end{figure}

We can also show that
\begin{align}
\partial_{m} \langle O \rangle = \lim_{x_c \to \infty}\Cov_{0,x_c} (O, \tr[ \tilde{D}^{-1}_{\rm ov}(m)])\,,
\end{align}
for sea-quark mass $m$ and
\begin{align}
\tilde{D}^{-1}_{\rm ov}(m) = \frac1{1-m}\left(D^{-1}_{\rm ov}(m) - 1 \right)\,,
\end{align}
with overlap operator $D_{\rm ov}$ \cite{Brower:2012vk,Blum:2014tka}.  We find that the traces of $\tilde{D}^{-1}_{\rm ov}(m)$
can be efficiently estimated using our tadpole field approach of Ref.~\cite{Blum:2015you}.
For domain-wall fermions, an additional flavor enters the path integral as the determinant ratio
\begin{align}
  \det(D(m) D^{-1}(1))
\end{align}
with five-dimensional Dirac operator $D(m)$.  For $m=1$ this factor is trivial and we can view including an additional flavor as changing the sea-quark mass down from $m=1$ to the target value.
In this way integrating the parametric derivative with respect to $m$ allows us to determine the effects of introducing an additional sea-charm quark.  Setting $O$ to the Wilson-flowed energy density,
allows us to determine the effect of the additional sea-charm quark to the Wilson-flow scales $t_0$
and $w_0$.  In Fig.~\ref{fig:mfpar}, we show the convergence as a function of $x_c$ for the $\beta$ derivative as well as the charm-quark mass derivative at $m=0.8$ of $E(t_f)$ with $t_f=2.01 \approx t_0/2$ on the 96I ensemble.  The lower scale $t_0/2$ allows for a statistically more precise estimate of the dependence of the lattice spacing on $\beta$ and the charm-quark mass.

\subsection{Finite-volume effects}\label{sec:fv}
In order to determine the finite-volume effects on $C(t)$, the
analysis groups explored two methods: a direct fit to the 24ID and 32ID data
as well as the Hansen-Patella approach \cite{Hansen:2019rbh,Hansen:2020whp}.  Details of the former
approach are given in Sec.~\ref{sec:distinctmethods}.  For the latter approach, we use
a monopole ansatz of the electromagnetic pion form factor
\begin{align}
F(k^2) = \frac{1}{1 - k^2 / m_\rho^2}
\end{align}
 and study the dependence on $m_\rho$.
 For this ansatz Ref.~\cite{Hansen:2020whp} gives an expression for the finite-volume corrections for $C(t)$ in terms of a simple integral
\begin{align}\label{eqn:hp}
C^{L}(t) - C^\infty(t)
   &=
   \sum_{ \vec{n} \neq \vec{0} }
   \frac{1}{ 6 \pi |\vec{n}| L }
   \Bigg\{
   {\rm Im}
   \int_{\mathbb{R}+i\mu} \frac{d k_3}{2\pi}
   \frac{ e^{i k_3 |x_0|} (4m_\pi^2 + k_3^2) m_\rho^4}{(m_\rho^2+k_3^2)^2}
   \frac{
   e^{-|\vec{n}| L \sqrt{m_\pi^2+\frac{k_3^2}{4}}}
   }{4 k_3} \notag\\
   &\qquad
   + \int \frac{d p_3}{2\pi}
   e^{-|\vec{n}| L \sqrt{m_\pi^2+p_3^2}}
   \frac{d}{dz} \left[
   \frac{ e^{- z |x_0|} (z^2-4m_\pi^2) m_\rho^4}{(z+m_\rho)^2 (z^2 + 4p_3^2)}
   \right]_{z = m_\rho}
   \Bigg\}
   \,,
\end{align}
where $C^L$ is the correlator at finite spatial volume $L^3$ and $C^\infty$ is the infinite-volume version.  The equation depends on the pion mass $m_\pi$ and the monopole-mass parameter $m_\rho$.
The complex shift $i\mu$ of the integration contour has to be chosen in the range $0<\mu< 2 m_\pi$, however, the integral does not depend on the exact choice.
Equation \eqref{eqn:hp} only considers the pole contribution to the Compton amplitude and neglects terms of order $e^{-\sqrt{2 + \sqrt{3}} m_\pi L}$ as well as effects of finite Euclidean time.
This is well justified for our current precision goal.  The effects of the regular contribution to the Compton amplitude and effects of the finite Euclidean time extent are known \cite{Hansen:2019rbh,Hansen:2020whp} and may be considered in future work.

Note that the finite-volume corrections for the quark-connected diagram are $\frac{10}{9}$ of the total as is easily seen from the following argument.  Consider a theory with quark charges $Q_u=\frac12=-Q_d$ instead of the physical $Q_u=\frac{2}{3}=-2Q_d$.  The QED charges of mesons made of up and down quarks are identical in both cases, however, in the $Q_u=\frac12=-Q_d$ theory the quark-disconnected diagram does not contribute, while the quark-connected diagram contributes with a $Q_u^2 + Q_d^2 = \frac12$ factor instead of the physical $Q_u^2 + Q_d^2 = \frac59$.  We therefore find that $\frac12 \frac95 = \frac{9}{10}$ of the quark-connected contribution is equal to the total contribution and equivalently that the total correction needs to be multiplied by $\frac{10}{9}$ to obtain the correction for the quark-connected piece.  This simple argument is consistent with partially quenched Chiral Perturbation Theory studies \cite{DellaMorte:2010aq,Aubin:2019usy,Lehner:2020crt}.

\section{Relative unblinding}\label{sec:relativeblind}
In the following, we summarize the different approaches of the five analysis groups and show the result of our relative unblinding process.  We highlight important findings and explain the prescription that all five groups agreed to be used for the full unblinding.

\subsection{Distinct methods of the five analysis groups}
\label{sec:distinctmethods}
Each analysis group received the blinded correlator data as described in Sec.~\ref{sec:blind}.  The separate analysis groups then discussed the data and agreed on the respective analysis methods within each group.  The confinement of these discussions to the separate groups lead to a diverse set of approaches to the data analysis.  In the following sub-sections, we briefly describe the approaches of each group, focusing on the differences.

\subsubsection{Group A}
Analysis group A provides results for $a_\mu^{\rm W}$ as well as $a_\mu^{\rm SD}$.  Statistical errors are obtained from a super-jackknife procedure \cite{CP-PACS:2001vqx,LHPC:2010jcs} for most ensembles combined with a binning study and using the master-field error estimates of Sec.~\ref{sec:mferr} on ensemble 96I.  The continuum extrapolations are performed based on the 24 data points over three lattice spacings described in Sec.~\ref{sec:overviewimpro}, where small linear corrections to shift the individual points to the lines of constant physics (LCP) are applied first.  Finite-volume corrections are also applied before the continuum extrapolation.
To this end, the Hansen-Patella Eq.~\eqref{eqn:hp} is used for finite-volume corrections with nominal parameters $m_\rho=727$ MeV and errors estimated from the variation to $m_\rho=770$ MeV.  An additional ad-hoc $20\%$ uncertainty is added to the finite-volume corrections to account for the limitations discussed in Sec.~\ref{sec:fv}.  Combinations of the fit ansaetze
\begin{align}
  f_{2}(a^2) &= c_0 + c_1 a^2 \,, \\
  f_{2,4}(a^2) &= c_0 + c_1 a^2 + c_2 a^4 \,, \\
  f_{2\alpha}(a^2) &= c_0 + c_1 a^2 \alpha_s(\mu=1/a) \,, \\
  f_{2\alpha,4}(a^2) &= c_0 + c_1 a^2 \alpha_s(\mu=1/a) + c_2 a^4
\end{align}
are then considered with four-loop running coupling $\alpha_s$ in the $\overline{\rm MS}$ scheme \cite{vanRitbergen:1997va}.

For $a_\mu^{\rm W}$, the central value is chosen as the average of the $f_2$ fits to the $(\omega_t,C^{\rm lc},Z_V^\star)$,
$(\omega_t,C^{\rm ll},Z_V^\star)$, $(\omega_t,C^{\rm lc},Z_V)$ trajectories
with $t^\star = 1$ fm.  These trajectories had the smallest $a^4$ contributions.  For $a_\mu^{\rm W}$, the effect of $\omega_t$ compared to $\hat{\omega}_t$ is negligible.
The continuum extrapolation error is estimated by varying $f_2$ to $f_{2\alpha}$ and by considering the spread of the mean to the individual $(\omega_t,C^{\rm lc},Z_V)$ and $(\omega_t,C^{\rm ll},Z_V^\star)$ fits.

For $a_\mu^{\rm SD}$, the fit form $f_{2,4}$ is used for all trajectories and the average of $(\hat{\omega}_t,C^{\rm lc},Z_V)$ and $(\hat{\omega}_t,C^{\rm lc},Z_V^\star)$ is used for the central value since they exhibit the smallest $a^4$ coefficients.  The variation from $f_{2,4}$ to $f_{2\alpha,4}$ as well as the maximal variation to $(\hat{\omega}_t,C^{\rm ll},Z_V)$, $(\omega_t,C^{\rm lc},Z_V)$,
$(\hat{\omega}_t,C^{\rm lc},Z_V)$, $(\hat{\omega}_t,C^{\rm ll},Z_V^\star)$, $(\omega_t,C^{\rm lc},Z_V^\star)$, and
$(\hat{\omega}_t,C^{\rm lc},Z_V^\star)$ is then used for the continuum extrapolation error.

The effects of the residual mass and the sea-charm quark are studied separately and found to be small compared to the quoted uncertainties.

\subsubsection{Group B}
Analysis group B provides results for $a_\mu^{\rm W}$ as well as $a_\mu^{\rm SD}$.
The strategy is to employ a global fit to all of the measurements on the ensembles listed in Sec.~\ref{tab:ex}. Statistical errors for each measurement, including lattice spacings, pion masses, and so on, are incorporated through a super-jackknife method. 

Several terms comprise the global fit function for the intermediate window. A second-order polynomial in $a^2$ is used to extrapolate non-zero lattice spacing to the continuum limit. Finite-volume effects are treated explicitly through a term exponential in $m_\pi L$ and are mainly constrained by the two Iwasaki-DSDR ensembles in Tab.~\ref{tab:ex}. Small light-quark-mass mistunings are treated linearly in the appropriate meson-mass squared and a simple linear ansatz for the residual mass is applied. Charm-quark mistunings are corrected with inverse mass-squared of the $D_s$ meson. All together, the fit function takes the form
\begin{align}
    a_\mu(...) &= a_\mu \left( 1+ c_1 a^2 + c_2 a^4 \right)\left( 1+ c_3 e^{-m_\pi L}\right)\left( 1+ c_4 (m_\pi^2-m^2_{\pi,\rm phys}) \right)\left(1+c_5 (m_K^2-m^2_{K,\rm phys})\right)\nonumber\\
    &\times\left(1+c_6 a m_{\rm res}\right)
    \left( 1+ c_7 \left(\frac{1}{m_{Ds}^2}-\frac{1}{m_{Ds,\rm phys}^2}\right) \right).
    \label{eq:group B global fit}
\end{align}
The coefficients $c_1$ and $c_2$ take on different values for the Iwasaki-DSDR ensembles, and the residual mass term is treated as an $O(a)$ artifact.

To fit the data to Eq.~\eqref{eq:group B global fit}, the (log of) $C(t)$ is first cubically interpolated between time-slices and then integrated with the continuum form of the one-loop QED kernel, Eq.~\eqref{eq:wt}. The central value of the procedure is determined from the average of conserved-local and local-local correlation functions for the HVP. The main part of the systematic error arises from the difference of these two results in the continuum limit.

For the short-distance window, the procedure is similar except that the discrete version of the one-loop kernel $\hat{\omega_t}$ is also used (approximated as $w_t(1-a^2/t^2)$) and an $a^2\log{a^2}$ term is considered. The systematic error is computed from differences between pairwise combinations of $a^2$, $a^4$ and $a^2\log{a^2}$ terms, using both $w_t$ and $\hat{w}_t$ weights, all added in quadrature. The central value is taken as the $w_t$ version with the conserved-local correlation function since empirically it has the smallest $a^4$ contamination. 

\subsubsection{Group C}
Analysis group C provides results for $a_\mu^{\rm W}$.
The strategy is divided in a few steps. First, using the ensembles listed in Tab.~\ref{tab:ex} the derivatives of the intermediate window with respect to the quark masses are calculated. 
Additional cutoff or finite-volume effects on the derivatives are neglected.  The derivatives are then used to shift the three reference ensembles, 48I, 64I and 96I, to the LCP. Additionally, all windows are shifted to $m_\pi L=4$ using Chiral Perturbation Theory and additional systematic effects are not considered since they are well below the statistical uncertainty.

After multiplying by the normalization factors $Z_V$ or $Z_V^\star$, the intermediate windows from the 3 ensembles and 2 discretizations ($C^{\rm ll}$ and $C^{\rm lc}$) are extrapolated to the continuum limit with a constrained fit. Note that also $a^2/t_0$ used in the extrapolation is shifted to the proper LCP. The following three types of fits are considered: linear and quadratic in $a^2$ with all 6 data points and linear in $a^2$ with the finest 4 data points (96I, 64I).
A systematic error from the spread of the central values of the fitted continuum windows is included in the error budget. Both correlated and uncorrelated fits are used, and for the latter their quality is assessed using the method developed in Ref~\cite{Bruno:2022mfy}. The 3 fits described above are performed separately using $Z_V$ and a variant of $Z_V^\star$. For the former it is observed that the linear fit in $a^2$ is not acceptable, and that a quadratic term is necessary to describe the data. Hence, the preferred strategy is based on $Z_V^\star$ and the preferred fit is the constrained linear fit to all 6 data points. For the variant of $Z_V^\star$, a slight modification of the definition provided in Sec.~\ref{sec:overviewimpro} is considered, i.e., the ratio of $C^{\rm lc}$ over $C^{\rm ll}$ is used individually integrated using the smearing function $\Theta(t,t^\star-\Delta/2,\Delta) \Theta(t^\star+\Delta/2,t,\Delta)$ with $\Delta=0.15~\mathrm{fm}$. Several values of $t^\star$ are explored and for the final analysis $t^\star=1~\mathrm{fm}$ is adopted. No particular difference is observed with respect to the interpolation described in Sec.~\ref{sec:overviewimpro}, as one can easily infer from the long plateau in Fig.~\ref{fig:ratiolcll}.

The statistical analysis is carried out by propagating all fluctuations of observables using both the Jackknife method and the $\Gamma$-method~\cite{Wolff:2003sm}. No large autocorrelations in the extrapolated continuum window are observed. Finite-volume effects to correct from $m_\pi L=4$ to $\infty$ are obtained from an independent implementation of Eq.~\eqref{eqn:hp}.
Final shifts for residual mass effects and dynamical charm effects are applied in the same manner as also done by group B.  

\subsubsection{Group D}
Analysis group D provides results for $a_\mu^{\rm W}$ from  
the physical pion-mass ensembles 48I, 64I, and 96I, which are computed
with a binned super-jackknife analysis with weight function $w_t$
and vector current normalizations $Z_V$ and $Z_V^\star$.
In addition, a version of $Z_V$ is used, where the pion state is replaced by a kaon state.
The mass extrapolation to the physical point is done by assuming linear dependence on the quark masses taken from ensemble 1 with 4 and ensemble 1 with 3, respectively.
Finite-$L_s$ effects are corrected by assuming linearity in $m_{\rm res}$
using ensembles 1, 2, 4, and A.  The values of
$a_\mu^{\rm W}$ on the 48I and 64I ensembles are corrected 
by an exponential dependence to the lattice extent, 
$\exp(-m_\pi L)$, 
whose coefficient is 
taken from the 24ID and 32ID ensembles,
to match for the volume of 96I.  A 50\%  systematic uncertainty for these finite-volume corrections is added.  It is noted that within the statistical noise of the 24ID and 32ID ensembles, their difference is reproduced by the Hansen-Patella finite-volume formula
as well as the Meyer-Lellouch-L\"uscher-Gounaris-Sakurai \cite{Meyer:2011um,Lellouch:2000pv,Gounaris:1968mw} approach.

After these corrections for 18 data points from 
three ensembles, two vector currents $C^{\rm ll}$ and $C^{\rm lc}$, and three vector current normalizations,
the continuum extrapolation is performed by combinations of
the fit formulae $f_{2}(a^2)$,  $f_{2,4}(a^2)$, $f_{2\alpha}(a^2)$,  and $f_{2\alpha,4}(a^2)$
by requiring a universal continuum limit for all 18 data points.
$f_{2}(a^2)$ poorly fits $C^{\rm ll}(t)$ with the coarsest ensemble 48I, 
and it is decided to drop this combination from the final results.
In analysis group D, the central value for the continuum extrapolation
is chosen from fit $f_{2}(a^2)$ to $C^{\rm ll}(t)$ and
$f_{2\alpha}(a^2)$ to $C^{\rm lc}(t)$. The error of the continuum extrapolation
is determined to cover all central values of the considered fit forms.
The continuum extrapolation for each of the 6 individual
combination of currents and normalizations is also performed.
The results are consistent with that of the universal fit except, again,
the $f_{2}(a^2)$ fit for $C^{\rm ll}(t)$.
Finally, a small volume correction from the 96I volume to infinity is carried out using
the Meyer-Lellouch-L\"uscher-Gounaris-Sakurai approach.
For each of the isospin-symmetric worlds, 
RBC/UKQCD18 and BMW20, the lattice spacing is determined in two different 
scaling trajectories (either keeping $w_0$ or $m_\Omega$ fixed).
The fit results are consistent between the two scaling trajectories,
providing an additional check for the continuum extrapolation of
$a_\mu^{\rm W}$.

\subsubsection{Group E}
Analysis group E provides results for $a_\mu^{\rm W}$.
The strategy is entirely data driven.
Statistical uncertainties are determined from a
bootstrap analysis with measurements within 20 MD units binned into an effective measurement.  The input uncertainties are propagated via re-sampling (Gaussian error propagation). 
Both $\omega_t$ and $\hat{\omega}_t$ kernels are used.
In addition to $Z_V$ a variant of $Z_V^\star$ is used
that for a given window is defined as
\begin{align}
 Z_V^{C} = \frac{a_\mu^{\rm lc,bare}}{a_\mu^{\rm ll,bare}}\,,
\end{align}
where $a_\mu^{\rm ab,bare}$ is obtained without vector-current normalization factors from the bare correlators $C^{ab}$.
When referring to $a_\mu^{Z,K}$ below, $a_\mu^{\rm ll,bare}$ is normalized using two powers of $Z_V$ or two powers of $Z_V^C$.
The chiral, strange-quark, discretization, and finite-volume effects are fitted to all ensembles for a given choice of
renormalization procedure and kernel to the ansatz
\begin{align}
        a_\mu^{Z,K} &= a_\mu^\mathrm{phys}\times \left(1 + C_\chi \frac{(m^2_\pi-{(m_\pi^2)}^\mathrm{phys})}{{(m_\pi^2)}^\mathrm{phys}} \right) \times \left(1 + C_s \frac{(X_s - X_s^\mathrm{phys})}{X_s^\mathrm{phys}} \right)\\
        & \qquad \quad \times \left(1 + C_V e^{-m_\pi L} \right)  \times  \left(1+C^{Z,K}_{CL,0} (a\Lambda)^2  + C^{Z,K}_{CL,1} (a\Lambda)^4\right)\times \left(1 + C^{Z,K}_5 am_\mathrm{res}\right) \,.
\end{align}
In this formula $X_s$ stands for $m_K$ for the RBC/UKQCD18 world and for $m_{ss\star}$ for the BMW20 world.
The ratios $R^{Z,K}_{Z',K'}$ on the three physical point Iwasaki ensembles are simultaneously fitted to the model $f_R$,
\begin{align}
        R^{Z,K}_{Z',K'} &\equiv \frac{a_\mu^{Z,K}}{a_\mu^{Z',K'}} \,,  & f_R &\equiv \frac{1+C^{Z,K}_{CL,0} (a\Lambda)^2  + C^{Z,K}_{CL,1} (a\Lambda)^4}{1+C^{Z',K'}_{CL,0} (a\Lambda)^2  + C^{Z',K'}_{CL,1} (a\Lambda)^4}
\end{align}
and the ratio $R^{\rm ID}_V$ for the ensembles 32ID and 24ID to the model $g_V$,
\begin{align}
        R^{\rm ID}_V &\equiv \frac{a_\mu^{\rm 32ID}}{a_\mu^{\rm 24ID}} \,, &        g_V &\equiv  \frac{1+C_V e^{-(m_\pi L)^{\rm 32ID}}}{1+C_V e^{-(m_\pi L)^{\rm 24ID}}} \,.
\end{align}
All correlations between data points on the same ensembles are included in this fit.
Systematic uncertainties are estimated by
variations on the data that enters the fit and/or the terms included in the model(s).

\subsection{Comparison of results}
After the analysis groups had individually converged on their respective methodology described
above, we started the process of relative unblinding.  The relative unblinding of groups $X$ and $Y$ was conducted by sharing the individually blinded data sets of group $X$ with group $Y$ and vice versa.  One of the groups then re-ran their analysis without modifications on the other data set.  This allowed for a direct comparison of groups $X$ to $Y$ while still keeping the absolute blinding intact.  

\begin{figure}
  \includegraphics[height=6cm,page=9]
  {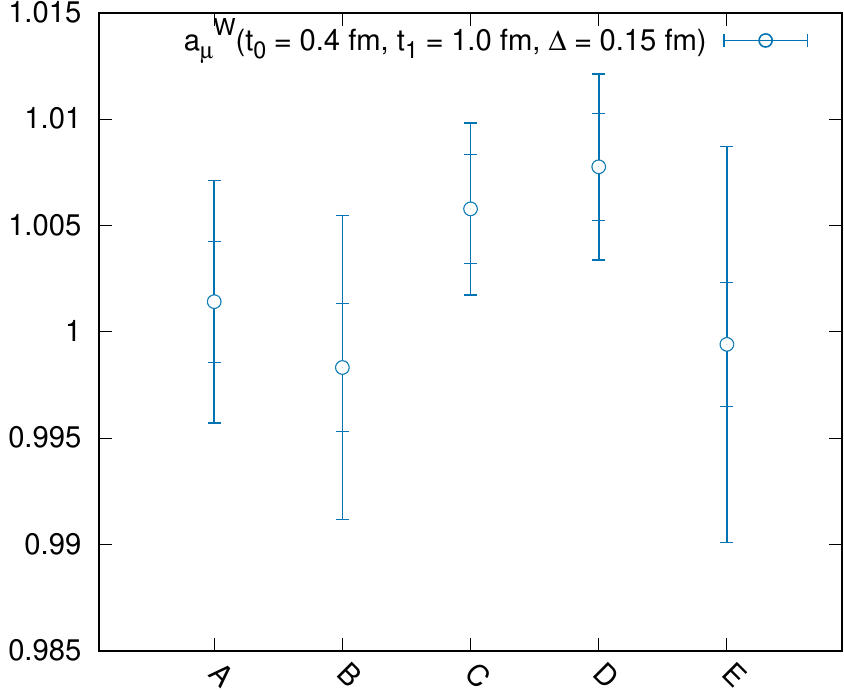}
  \hspace{0.4cm}
  \includegraphics[height=6cm,page=3]
  {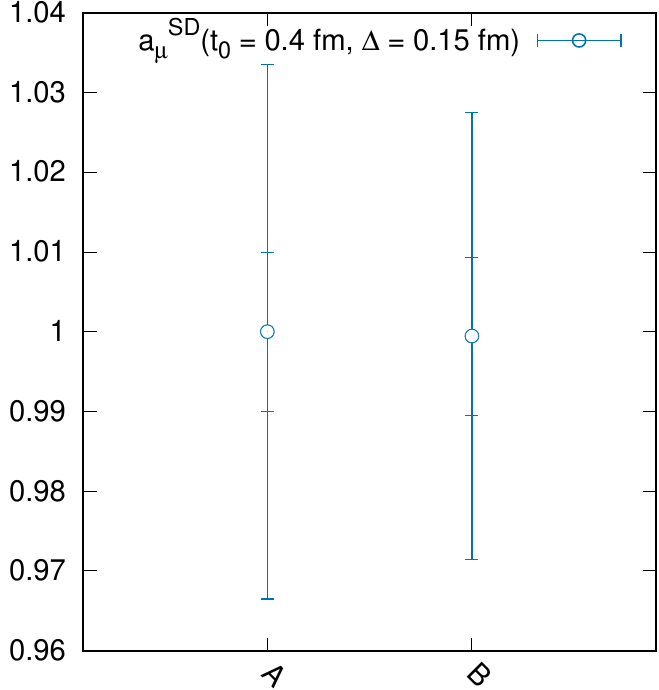}
\caption{\label{fig:relunblind} Result of the relative unblinding procedure for $a_\mu^{\rm W}$ (left) and $a_\mu^{\rm SD}$ (right).  The results are normalized to the preferred prescription described in Sec.~\ref{sec:preferred}.  The inner error bars show the statistical uncertainty, the outer error bars show the statistical and systematic uncertainties added in quadrature.}
\end{figure}

In Fig.~\ref{fig:relunblind}, we show the final result of the relative unblinding procedure for $a_\mu^{\rm W}$, for which all five groups participated.  The inner error bars give the statistical uncertainty, the outer error bars give statistical and systematic uncertainties added in quadrature.  We first note that the statistical uncertainties quoted by the separate analysis groups are consistent.  In addition, the different systematic approaches described in Sec.~\ref{sec:distinctmethods} yield different systematic uncertainties, however, all results are consistent within total uncertainties.

The blinding procedure described in Sec.~\ref{sec:blind} allows the $a^4$ term to affect the comparison at the level of $\pm 0.0025$ if the $a^4$ terms are not included in the fits.  This effect is small compared to the quoted uncertainties and is completely eliminated in Sec.~\ref{sec:unblind}, where we show the results of all groups after they repeated their unmodified analysis with the fully unblinded data sets.

  \subsection{Important findings}\label{sec:findings}
  After the relative unblinding process, the analysis groups exchanged their most important findings for our data sets.  We discuss these findings in this sub-section.  They form the basis, determined entirely on blinded data, of formulating the preferred prescription to produce the combined collaboration result described in Sec.~\ref{sec:preferred}.

      \begin{figure}
  \includegraphics[width=7cm,page=1]{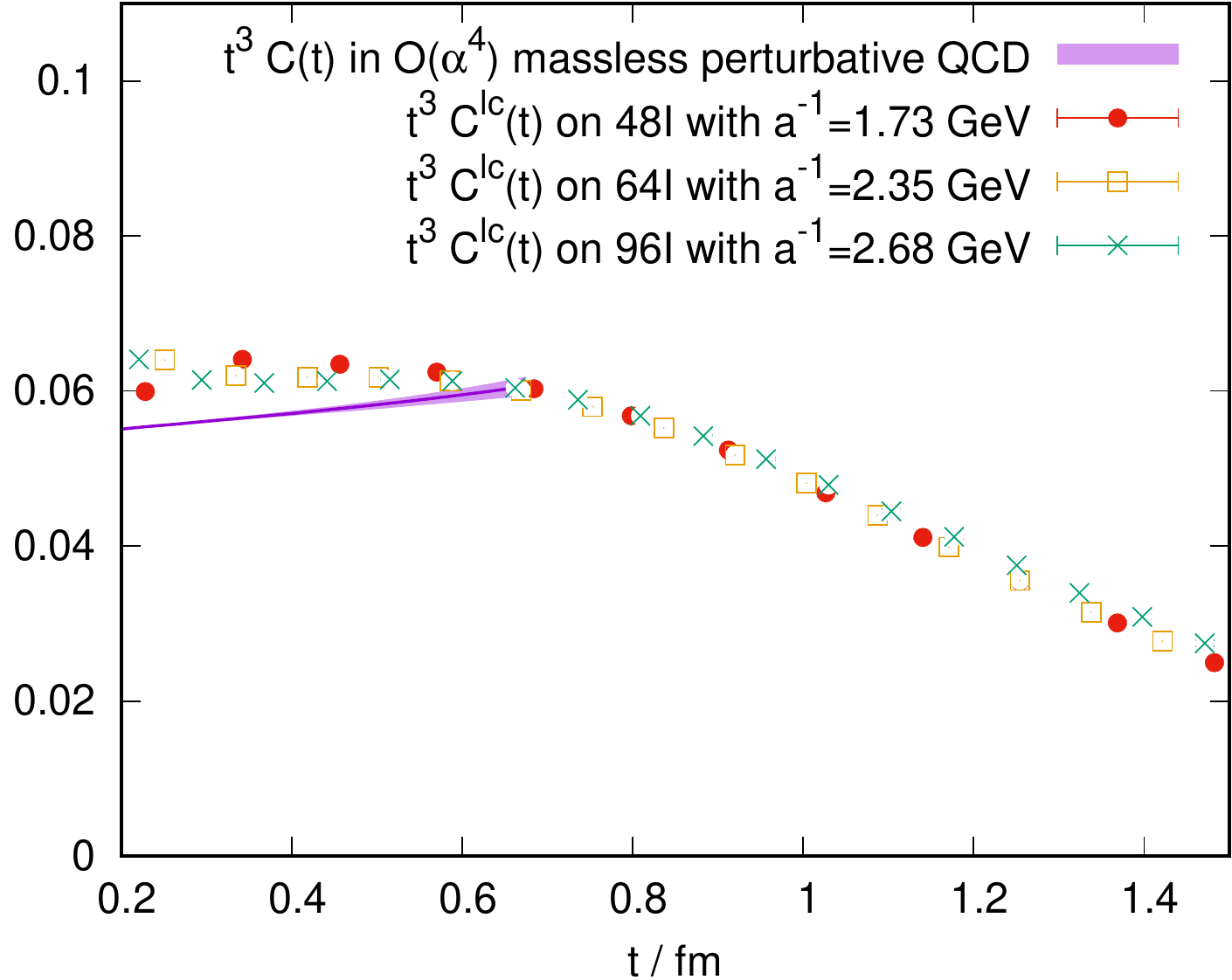}\hspace{1cm}\includegraphics[width=7cm,page=2]{figs/c3-crop}
\caption{\label{fig:ct3}The dimensionless correlation function combinations $t^3 C^{\rm lc}(t)$ (left) and $t^3 C^{\rm ll}(t)$ (right) as well as the perturbative result obtained from Ref.~\cite{Chetyrkin:2010dx}.}
      \end{figure}
      
  \begin{description}
  \item[Finding 1] The correlator $C^{\rm ll}$ has significantly larger $a^2 / t^2$ and $a^4/t^4$ errors compared to $C^{\rm lc}$.  These errors also noticeably affect $a_\mu^{\rm W}$.
  In Fig.~\ref{fig:ct3}, we plot the dimensionless $t^3 C(t)$ to highlight this effect.

\item[Finding 2] Mean-field improved lattice perturbation theory finds the discretization errors of $C^{\rm ll}$ to be approximately double the discretization errors of $C^{\rm lc}$.

\item[Finding 3] When analyzing $a_
\mu^{\rm SD}$, where both $a^2$ and $a^4$ coefficients were determined, the size of the $a^4$ coefficient is substantially larger for $C^{\rm ll}$ compared to $C^{\rm lc}$.

\item[Finding 4] The continuum extrapolation is sensitive to how finite-volume corrections are applied to the individual ensembles.  This is an important effect in our analyses since the new finest 96I ensemble has a larger physical volume compared to the 64I and 48I ensembles.
  \end{description}

  \subsection{Preferred prescription}\label{sec:preferred}
Based on the findings outlined in Sec.~\ref{sec:findings}, the collaboration decided on the following principles for the combined analysis that will be used for the full unblinding.  First, when using $C^{\rm ll}$, we always add a $a^4$ term to the fits.  Second, we use the Hansen-Patella finite-volume corrections instead of the data-driven fits to $e^{-m_\pi L}$ since we expect the Hansen-Patella formalism to more precisely map out the volume dependence.

\begin{figure}
  \includegraphics[height=6cm,page=7]{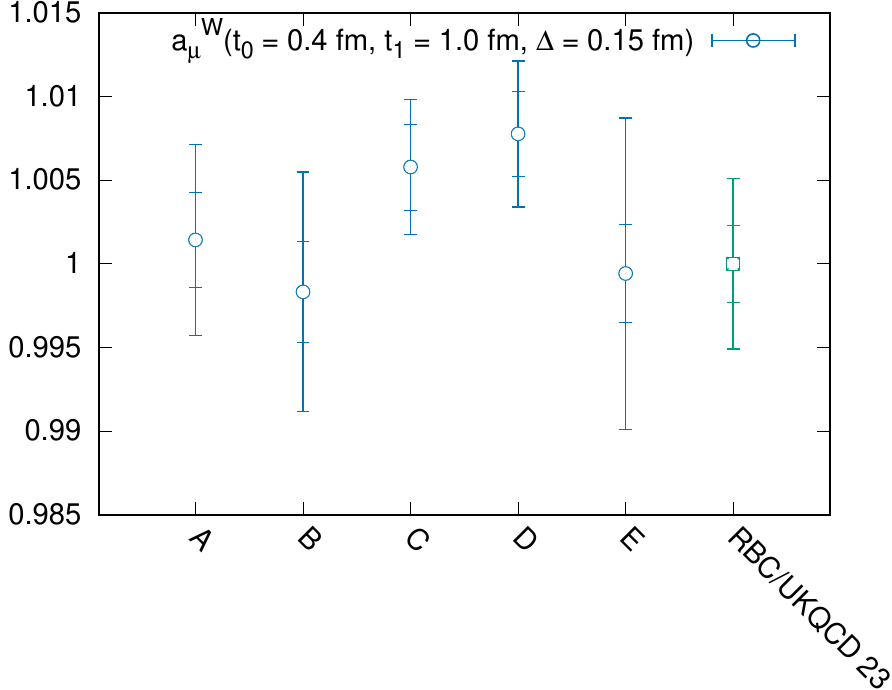}
  \caption{\label{fig:relunblindpref} Result of the relative unblinding procedure for $a_\mu^{\rm W}$ inlcuding the preferred prescription RBC/UKQCD 23 described in Sec.~\ref{sec:preferred}.  The data is normalized to the RBC/UKQCD 23 prescription.  The inner error bars show the statistical uncertainty, the outer error bars show the statistical and systematic uncertainties added in quadrature.}
\end{figure}

These principles are then implemented in the following prescription for $a_\mu^{\rm W}$.
For the vector current renormalization factor, we use $Z_V$ as well as $Z_V^\star$ with $t^\star=1$ fm.
For the weight functions we use $\hat{w}_t$ as well as $w_t$.  For the continuum extrapolation, we perform a simultaneous fit to the $C^{\rm ll}$ and $C^{\rm lc}$ data sets using
\begin{align}
 f_{\rm ll}(a^2) &= c_0 + c_1 a^2 + c_2 a^4 \,, \\
 f_{\rm lc}(a^2) &= c_0 + c_3 a^2
\end{align}
as well as
\begin{align}
 f_{{\rm ll},\alpha}(a^2) &= c_0 + c_1 a^2 \alpha_s(\mu=1/a) + c_2 a^4 \,, \\
 f_{{\rm lc},\alpha}(a^2) &= c_0 + c_3 a^2 \alpha_s(\mu=1/a) \,.
\end{align}
We therefore perform 8 fits in total.
We take the average of the minimum and maximum result as the central value for our prediction.  We take the difference of the central value to the maximum as our systematic error for the continuum extrapolation.  In Fig.~\ref{fig:relunblindpref}, we show the final result of the relative unblinding for each group as well as the preferred prescription, labelled RBC/UKQCD 23.  For $a_\mu^{\rm SD}$ the results of groups A and B were close to identical and we adopt the prescription of group A as the preferred result.

\section{Absolute unblinding}\label{sec:unblind}
After the collaboration converged on the preferred prescription described in Sec.~\ref{sec:preferred}, the analysis was frozen and the absolute unblinding was performed.  To this end, the unblinded data sets were distributed to the analysis groups, who then re-ran their analysis without modifications.  The results were presented by our collaboration already at the Edinburgh workshop of the g-2 Theory Initiative \cite{EdinburghTalk} in 2022 and are stated without modifications in the following.

\subsection{Intermediate-distance window $a_\mu^{\rm W}$}\label{sec:windowresult}
\begin{figure}
\includegraphics[width=10cm,page=1]{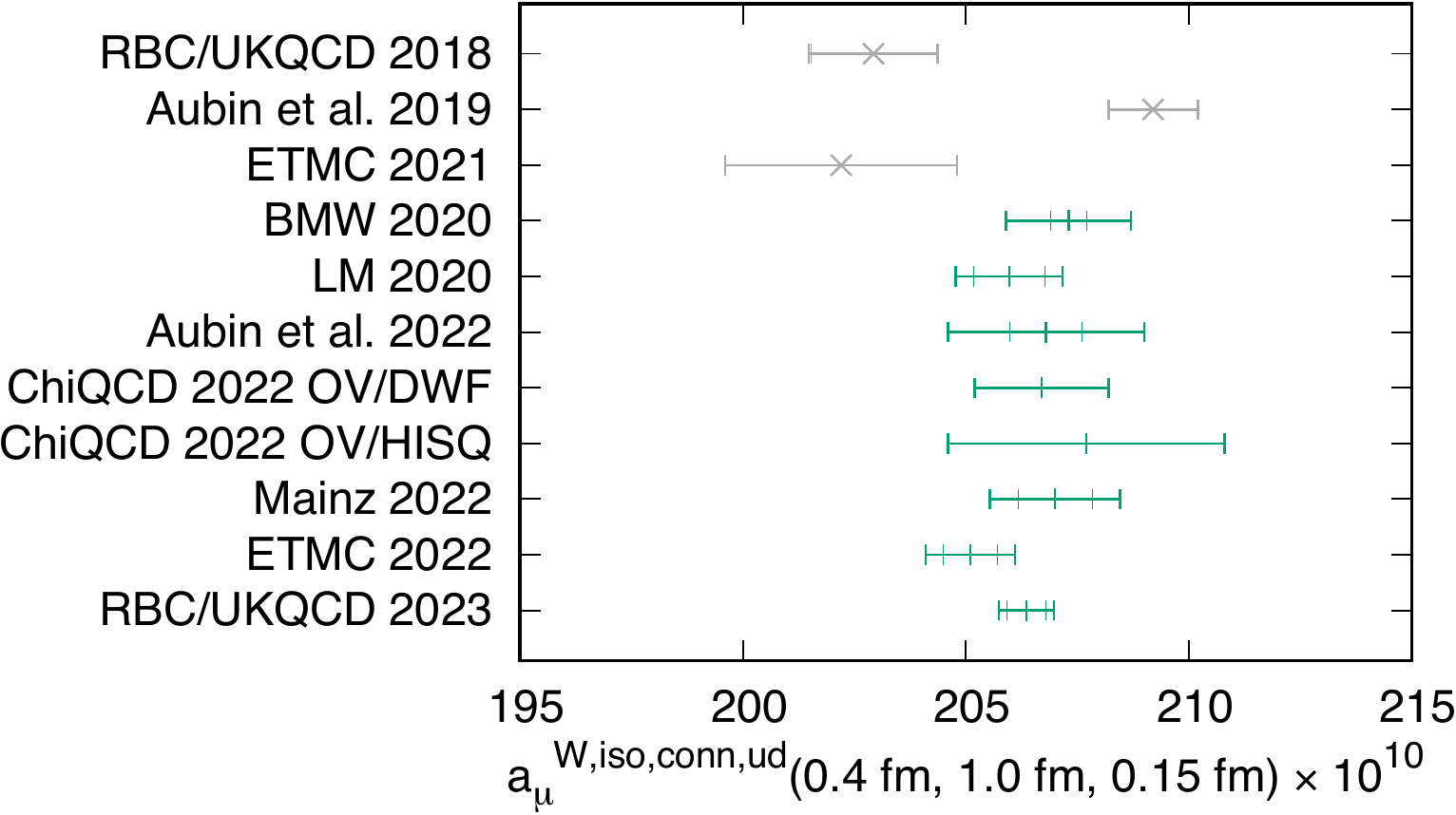}
  \caption{\label{fig:window-ud-conn-iso-comparison} Comparison of the up and down quark, connected, isospin-symmetric contribution to the intermediate window.  For historical completeness, we also show results that are superseded by newer results of the same collaboration at the top in gray.  The inner error bars show the statistical uncertainty, the outer error bars show the statistical and systematic uncertainties added in quadrature.
  RBC/UKQCD 2018 \cite{RBC:2018dos}, Aubin et al. 2019 \cite{Aubin:2019usy}, ETMC 2021 \cite{Giusti:2021dvd}, BMW 2020 \cite{Borsanyi:2020mff}, LM 2020 \cite{Lehner:2020crt}, Aubin et al. 2022 \cite{Aubin:2022hgm}, $\chi$QCD 2022 \cite{Wang:2022lkq}, Mainz 2022 \cite{Ce:2022kxy}, ETMC 2022 \cite{Alexandrou:2022amy}.
}
\end{figure}

For the intermediate-distance window $a_\mu^{\rm W}$ in the isospin-symmetric limit with $t_0=0.4$ fm, $t_1=1.0$ fm, and $\Delta=0.15$ fm, we find the up and down quark-connected contribution to
be
\begin{align}\label{eqn:intermediateconisolightbmwworld}
    a_\mu^{\rm W, iso, conn, ud} &= 206.36(44)_{\rm S}(42)_{\rm C}(01)_{\rm FV}(00)_{m_\pi~\rm FV}(08)_{\partial_m~\rm C}(00)_{\rm WF~order}(03)_{m_{\rm res}} \times 10^{-10}
\end{align}
in the BMW20 world and
\begin{align}\label{eqn:intermediateconisolightrbcworld}
    a_\mu^{\rm W, iso, conn, ud} &=206.46(53)_{\rm S}(43)_{\rm C}(01)_{\rm FV}(01)_{m_\pi~\rm FV}(09)_{\partial_m~\rm C}(00)_{\rm WF~order}(03)_{m_{\rm res}} \times 10^{-10}
  \end{align}
in the RBC/UKQCD18 world.
We separately quote the statistical uncertainties (S), the continuum limit uncertainties (C), the finite-volume uncertainties for the vector correlators (FV), the finite-volume uncertainties of the measured pion masses ($m_\pi$ FV), the uncertainties associated with the linear corrections to the line of constant physics ($\partial_m$ C), the uncertainties from the discretization of the Wilson flow equation (WF order), as well as the uncertainties due to the non-zero chiral symmetry breaking ($m_{\rm res}$).  The uncertainties from the ensemble-parameter and renormalization-factor determinations are fully propagated in the quoted uncertainties.  In Fig.~\ref{fig:window-ud-conn-iso-comparison}, we compare Eq.~\eqref{eqn:intermediateconisolightbmwworld} with previously published results.
In this work, we consistently use the BMW20 world for comparison plots of isospin-symmetric contributions.

\begin{figure}
\includegraphics[width=7.5cm,page=1]{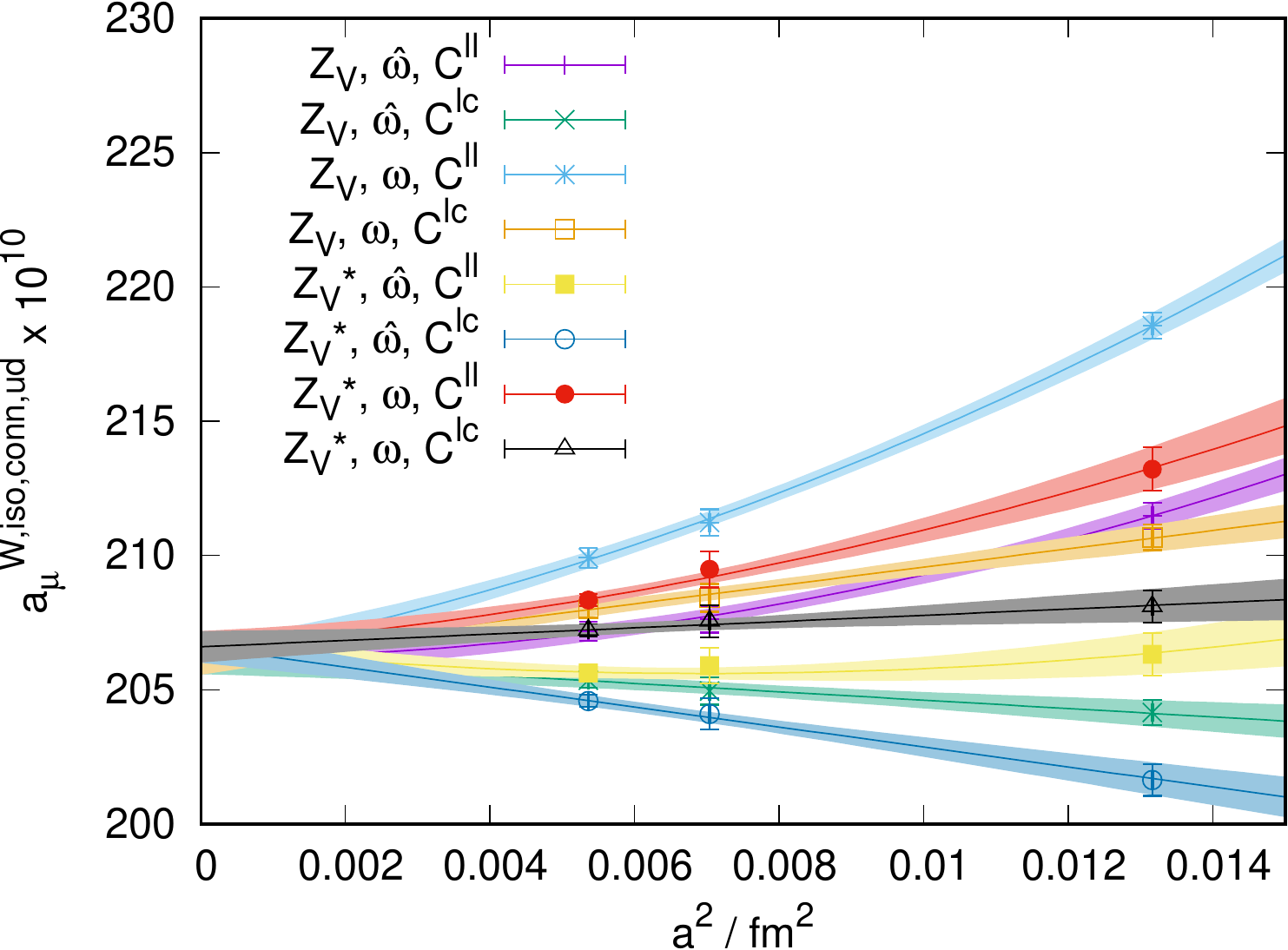}  \hspace{1.0cm}    \includegraphics[width=7.5cm,page=2]{figs/unblind-main}
  \caption{\label{fig:contextr} Continuum extrapolation of $a_\mu^{\rm W,iso,conn,ud} \times 10^{10}$.  On the left, we show the 8 fits of our preferred prescription.  On the right, we show the fit through the two data points already available in Ref.~\cite{RBC:2018dos} with lower statistical precision.}
\end{figure}

Compared to our earlier result presented in Ref.~\cite{RBC:2018dos}, where $a_\mu^{\rm W}$ was defined and computed for the first time, we increase the basis for our continuum extrapolation from 2 data points over two lattice spacings to 24 data points over three lattice spacings.  If we were to repeat the continuum extrapolation through the 2 data points already available in Ref.~\cite{RBC:2018dos} with lower statistical precision, we obtain a result consistent with the earlier work of $a_\mu^{\rm W, iso, conn, ud}=202.9(1.4) \times 10^{-10}$.  This is shown in Fig.~\ref{fig:contextr}.  The approximate 2$\sigma$ upward shift compared to Ref.~\cite{RBC:2018dos} can therefore dominantly be attributed to our improved continuum extrapolation.

\begin{figure}
\includegraphics[width=11cm,page=1]{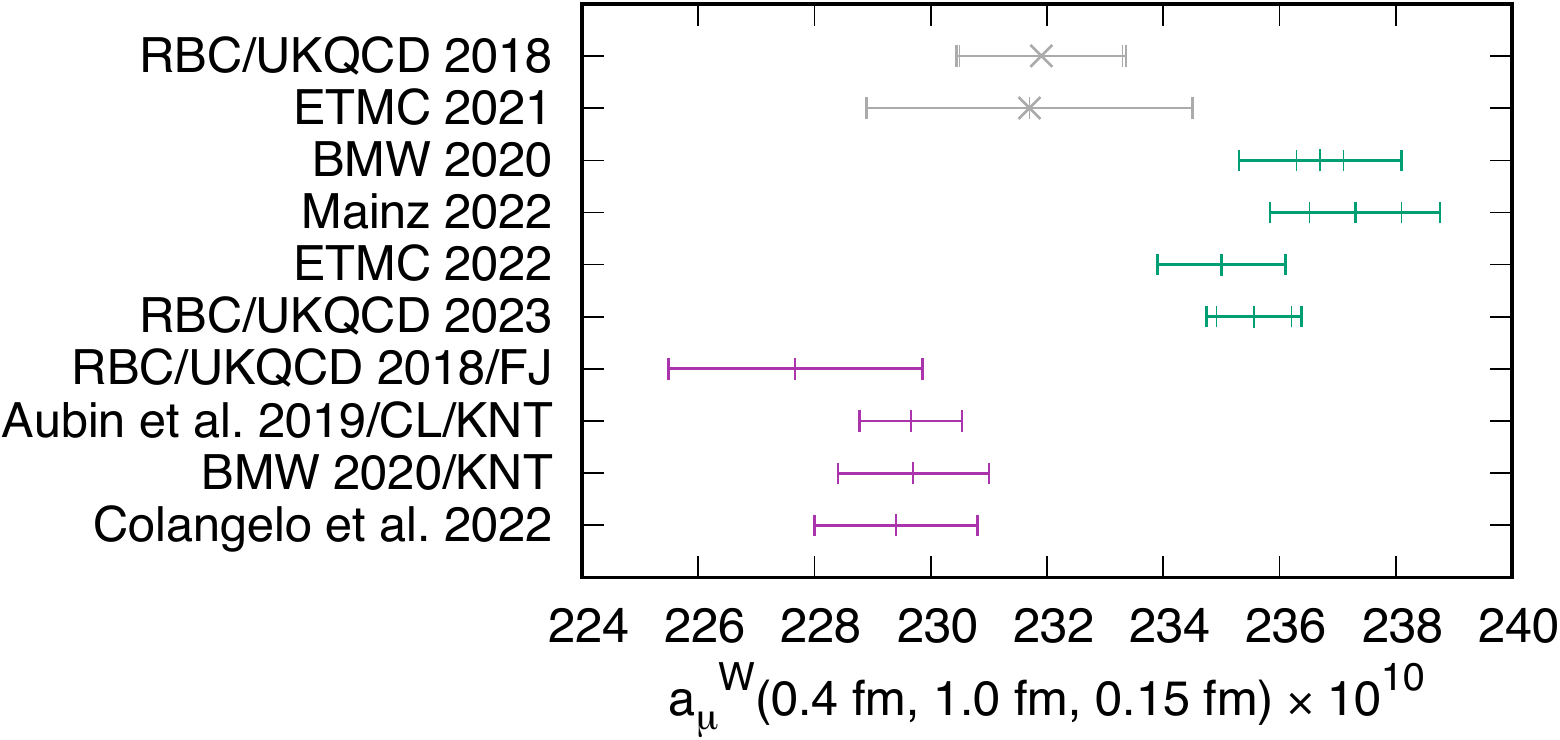}    
  \caption{\label{fig:window-comparison} Comparison of the total intermediate window contribution.  For historical completeness, we also show results that are superseded by newer results of the same collaboration at the top in gray.  Dispersive resuls are shown in purple, lattice results are shown in green.
  The inner error bars show the statistical uncertainty, the outer error bars show the statistical and systematic uncertainties added in quadrature.  RBC/UKQCD 2018 \cite{RBC:2018dos}, ETMC 2021 \cite{Giusti:2021dvd}, BMW 2020 \cite{Borsanyi:2020mff}, Mainz 2022 \cite{Ce:2022kxy}, ETMC 2022 \cite{Alexandrou:2022amy}, RBC/UKQCD 2018/FJ \cite{RBC2018FJ}, Aubin et al.~2019/CL/KNT \cite{Aubin2019CLKNT}, BMW 2020/KNT \cite{BMW2020KNT}, Colangelo et al. 2022 \cite{Colangelo:2022vok}.}
\end{figure}

In Ref.~\cite{RBC:2018dos}, we also computed the QED, strong-isospin-breaking, strange, charm, and quark-disconnected contributions to the intermediate window quantity.  These contributions are much smaller in magnitude and their uncertainties due to the continuum extrapolation are much smaller in absolute terms compared to $a_\mu^{\rm W, iso, conn, ud}$.   By combining these contributions with our improved light quark-connected, isospin-symmetric result of Eq.~\eqref{eqn:intermediateconisolightrbcworld}, we obtain our prediction
for the total intermediate window contribution
\begin{align}
   a_\mu^{\rm W} = 235.56(65)(50) \times 10^{-10}
\end{align}
with statistical (left) and systematic (right) errors given separately.
This can be compared with other lattice results as well as results based on the R-ratio, see Fig.~\ref{fig:window-comparison}.
Our result is in $3.8\sigma$ tension with the recently published dispersive result of $a_\mu^{\rm W} = 229.4(1.4) \times 10^{-10}$ \cite{Colangelo:2022vok} and in agreement with recent lattice results \cite{Borsanyi:2020mff,Ce:2022kxy,Alexandrou:2022amy}.

\subsection{Short-distance window $a_\mu^{\rm SD}$}
For the short-distance window $a_\mu^{\rm SD}$ in the isospin-symmetric limit with $t_0=0.4$ fm and $\Delta=0.15$ fm, we find the up and down quark-connected contribution to be
\begin{align}\label{eqn:sdpurebmw}
 a_\mu^{\rm SD, iso, conn, ud} = 48.7(0.5)(1.6) \times 10^{-10}
\end{align}
in the BMW20 world
and
\begin{align}\label{eqn:sdpurerbc}
 a_\mu^{\rm SD, iso, conn, ud} = 49.0(0.6)(1.4) \times 10^{-10}
\end{align}
in the RBC/UKQCD18 world.  We can substantially improve this result by replacing
the very shortest distances with perturbative QCD.
Such a hybrid result of perturbative and non-perturbative QCD is still a first-principles
determination but may combine the strength of both approaches.  In addition, the study of the consistency of lattice QCD and perturbative QCD at short distances may play an important role in understanding the origin of the tension for $a_\mu^{\rm W}$ described in Sec.~\ref{sec:windowresult}.

\begin{figure}
  \includegraphics[width=10cm]{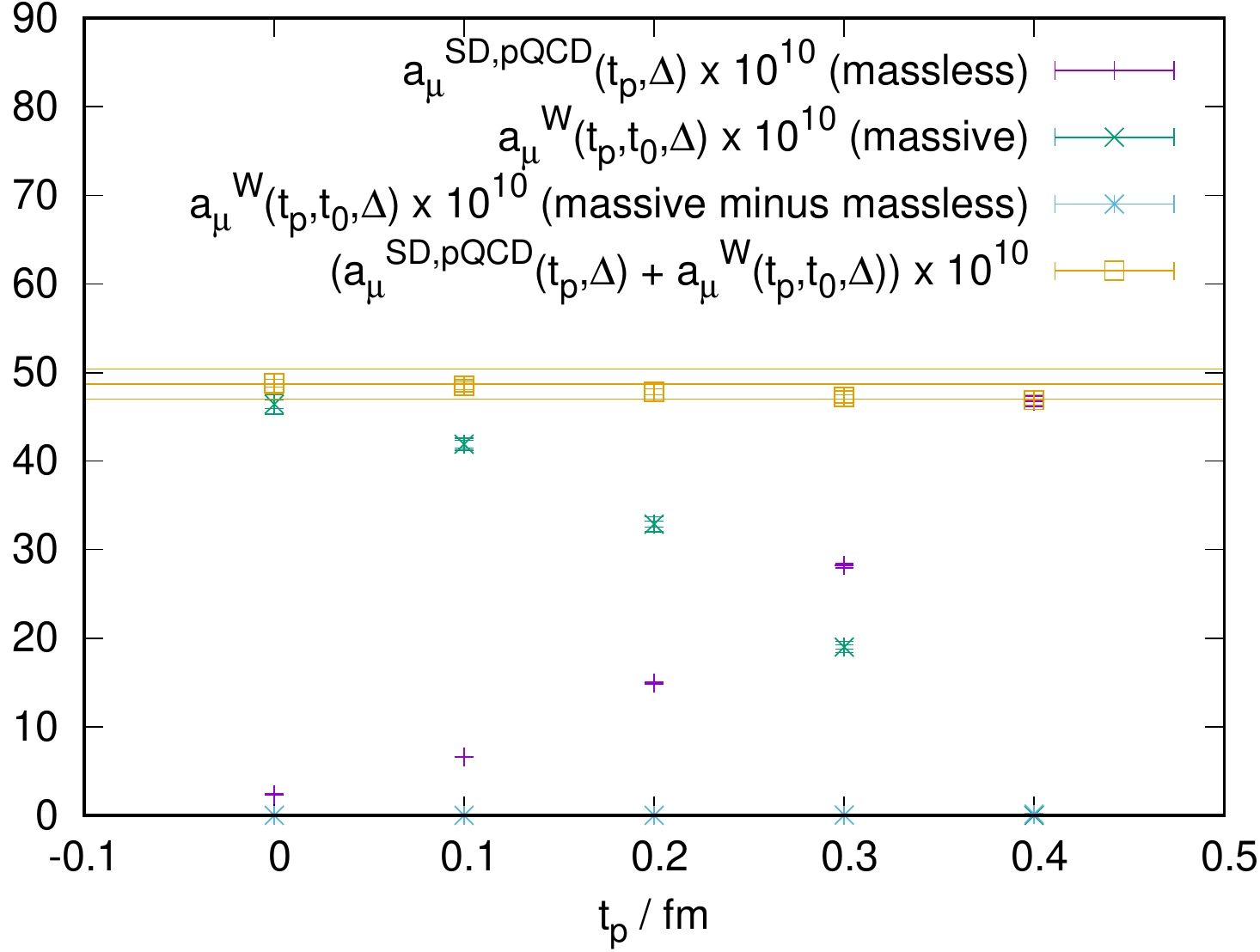}
  \caption{\label{fig:stabilitySD} Stability plot of Eq.~\eqref{eqn:stab} for $t_0=0.4$ fm and $\Delta=0.15$ fm.  The massless perturbative QCD result is taken from Ref.~\cite{Chetyrkin:2010dx}.  The correction from zero quark mass to non-zero quark mass is obtained from a linear extrapolation in the quark mass using ensembles 48I, 1, and 4.  The horizontal lines give the result of lattice QCD without combination with perturbative QCD.  Only the quark-connected isospin-symmetric up and down quark contribution is shown.
  }
\end{figure}

\begin{figure}
\includegraphics[width=10cm,page=1]{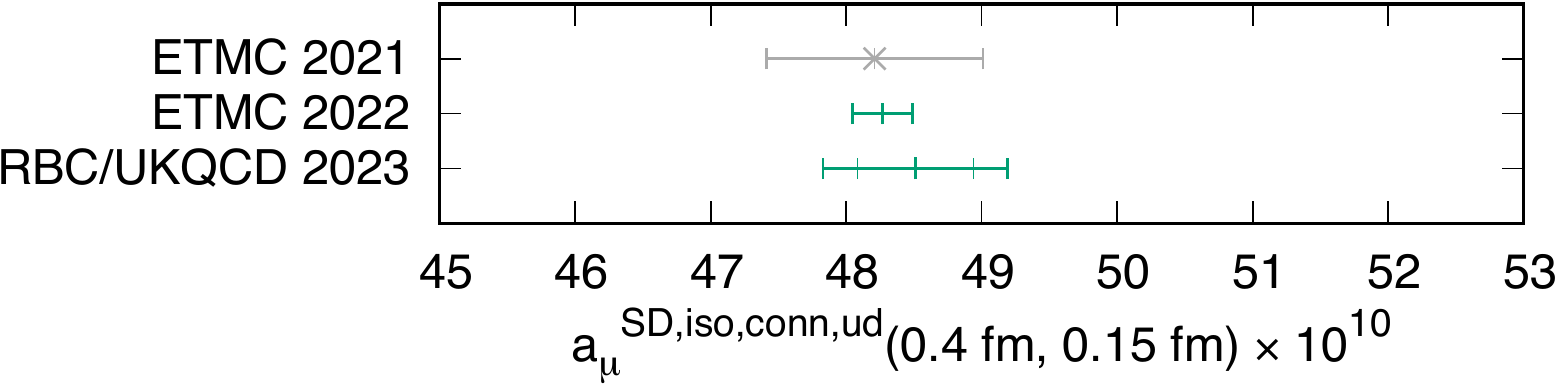}     
  \caption{\label{fig:window-ud-conn-iso-sd-comparison} Comparison of our preferred result with previous determinations.  For historical completeness, we also show results that are superseded by newer results of the same collaboration at the top in gray.
 The inner error bars show the statistical uncertainty, the outer error bars show the statistical and systematic uncertainties added in quadrature.  ETMC 2021 \cite{Giusti:2021dvd}, ETMC 2022 \cite{Alexandrou:2022amy}.
  }
\end{figure}

To establish a hybrid method, we use the additive property of the windows, i.e.,
\begin{align}
a_\mu^{\rm SD}(t_0,\Delta) = a_\mu^{\rm SD}(t_p,\Delta) + a_\mu^{\rm W}(t_p,t_0,\Delta) \,.
\end{align}
We can then evaluate the first term in perturbative QCD at $O(\alpha^4)$~\cite{Chetyrkin:2010dx} and the second term in lattice QCD, i.e., we write
\begin{align}\label{eqn:stab}
a_\mu^{\rm SD}(t_0,\Delta) = a_\mu^{\rm SD, pQCD}(t_p,\Delta) + a_\mu^{\rm W}(t_p, t_0, \Delta) \,.
\end{align}
In Fig.~\ref{fig:stabilitySD}, we study this separation as a function of $t_p$.  To the degree
that perturbative QCD agrees with lattice QCD at distance $t_p$, the plot should exhibit a plateau.
We find that lattice QCD and perturbative QCD are consistent within $1.5 \times 10^{-10}$ up to $0.4$ fm.  For a related study of matching perturbative QCD to short-distance vector current correlators, see Ref.~\cite{Giusti:2018mdh}.  If we choose $t_p=0.1$ fm, we find
\begin{align}\label{eqn:bmwworldsd}
  a_\mu^{\rm SD, iso, conn, ud} = 48.51(43)(53) \times 10^{-10}
\end{align}
in the BMW20 world and
\begin{align}
  a_\mu^{\rm SD, iso, conn, ud} = 48.70(52)(59) \times 10^{-10}
\end{align}
in the RBC/UKQCD18 world.  This is our preferred prescription for $a_\mu^{\rm SD, iso, conn, ud}$.  We compare Eq.~\eqref{eqn:bmwworldsd} to previous results in Fig.~\ref{fig:window-ud-conn-iso-sd-comparison}.  The hybrid method reduces the large discretization errors for the short-distance window and specifically also reduces the logarithmic discretization errors described in Refs.~\cite{Ce:2021xgd} and \cite{Chimirri:2022gzg}.

\begin{figure}
  \includegraphics[width=8cm]{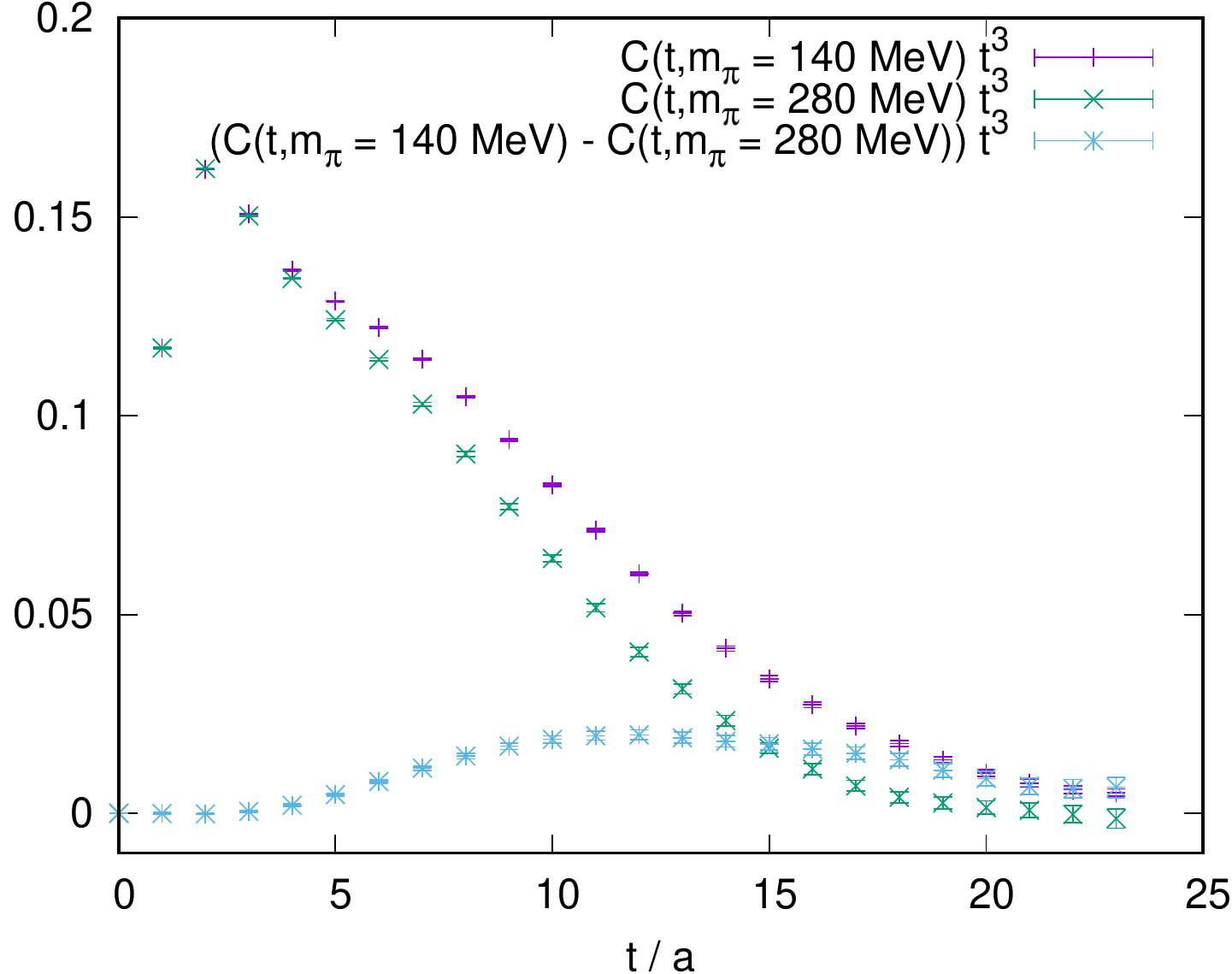}
  \caption{\label{fig:massdependence} Mass dependence of the vector correlator on a lattice with $a^{-1}=1.73$ GeV.  At very short distances, the vector correlator is effectively independent of the quark mass.}
\end{figure}

Finally, we note that the short-distance correlator is insensitive to the quark mass, see Fig.~\ref{fig:massdependence}.
This motivates a new approach to study the continuum limit of the HVP.  Since discretization errors largely cancel in the difference between vector currents evaluated at different quark masses, we proposed a mass-splitting approach in Ref.~\cite{snowmass2021loi}.  In this approach, we generate pairs of ensembles with $m_\pi$ and $M_\pi$ with $M_\pi \gg m_\pi$ to compute
\begin{align}
  a_\mu(m_\pi) = \underbrace{a_\mu(m_\pi) - a_\mu(M_\pi)}_{\equiv \delta a_\mu} + a_\mu(M_\pi) \,.
\end{align}
This allows us to consider the continuum limit of $\delta a_\mu$ and $a_\mu(M_\pi)$ separately.  The costly term $\delta a_\mu$ can then be calculated at coarser lattice spacings compared to $a_\mu(M_\pi)$.  This method will be used in upcoming improvements to the present calculation.

\subsection{Isospin-symmetric scheme dependence}
For comparisons of quantities defined in an isospin-symmetric world, it is crucial to precisely match the definitions of the isospin-symmetric point.  In Sec.~\ref{sec:iso}, we defined two hadronic schemes to define the isospin-symmetric world that match results previously presented by the RBC/UKQCD and BMW collaborations.  In previous sections, we presented our results separately for both schemes.  In this section, we provide results for the correlated difference of the BMW20 minus the RBC/UKQCD18 world.
For the intermediate window we find
\begin{align}
  \Delta a_\mu^{\rm W, iso, conn, ud} = -0.10(24)(07) \times 10^{-10}
\end{align}
and for the short-distance window we find
\begin{align}
  \Delta a_\mu^{\rm SD, iso, conn, ud} = -0.33(36)(36) \times 10^{-10}
\end{align}
using the lattice results of Eqs.~\eqref{eqn:sdpurebmw} and \eqref{eqn:sdpurerbc}.  We can therefore not
yet resolve the difference in isospin-symmetric schemes and they can be viewed as compatible at the current precision.

\subsection{Retrospective discussion of the blinding procedure}

\begin{figure}
  \includegraphics[width=7cm,page=7]{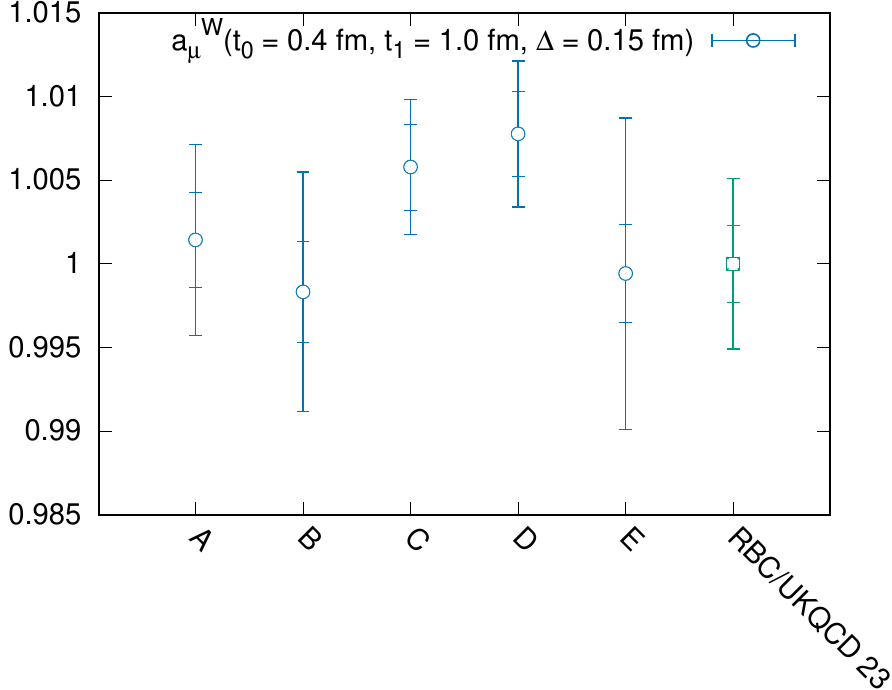}
  \includegraphics[width=7cm,page=8]{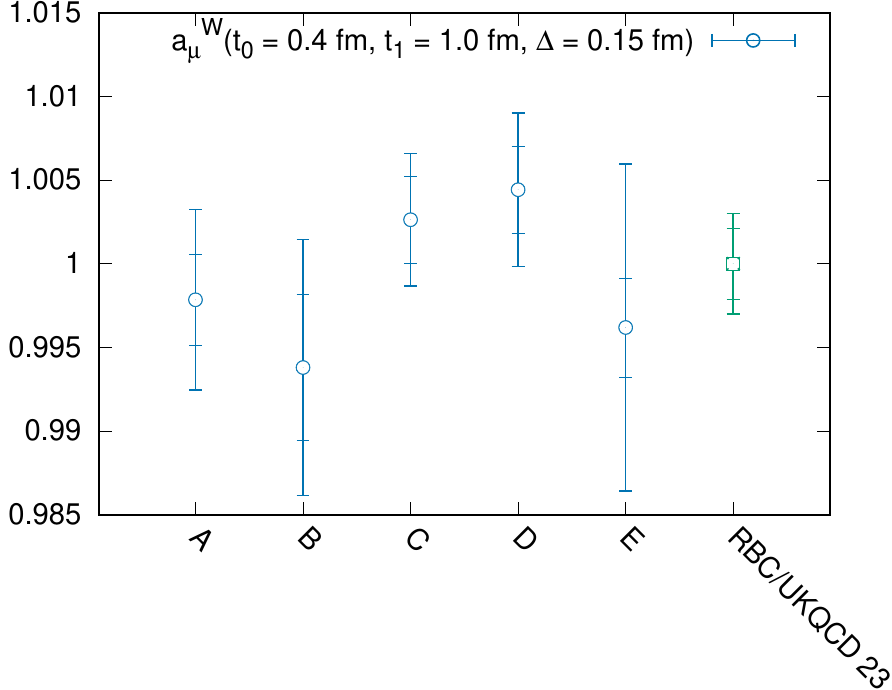}
  \caption{\label{fig:effectblinding} We show the result of the relative unblinding for $a_\mu^{\rm W}$ including the preferred prescription.  On the left side, each group used its own blinded data set including the $a^2$ and $a^4$ terms added in Eq.~\eqref{eqn:blind}.  On the right side, each group re-ran their unmodified analysis after the absolute unblinding on the unblinded dataset.  As anticipated, the artificial discretization errors in the blinded data can change central values and error estimates at the $\pm 0.0025$ level.  The data is normalized to the RBC/UKQCD 23 prescription.  The inner error bars show the statistical uncertainty, the outer error bars show the statistical and systematic uncertainties added in quadrature.}
\end{figure}

In the current paper, we performed a blinded analysis as described in Sec.~\ref{sec:blind}.  The goal of this procedure was to eliminate psychological bias that may have influenced systematic decisions of the analysis groups to favor either a larger value for $a_\mu^{\rm W}$, confirming the lattice QCD result of the BMW collaboration for this window quantity, or a smaller value, confirming the result based on the R-ratio.
To this end, we added artificial discretization errors using both $a^2$ and $a^4$ terms such that it is impossible for those who had access to our previous results for the coarser two lattice spacings of Ref.~\cite{RBC:2018dos} to completely unblind themselves by comparing the new blinded correlators with the previously shared data.  This is the reason for the three parameters of Eq.~\eqref{eqn:blind} exceeding the number of previously available lattice spacings.

Nevertheless, the possibility of an analysis group computing unblinded correlators based on the used gauge fields always remains.  Given the reduced statistical noise of short-distance time-slices of $C(t)$, even our chosen blinding procedure can in principle be circumvented with sufficient effort.
It therefore remains an important task to evaluate the balance between the threshold preventing such unblinding and the possible drawbacks introduced by the blinding procedure.
We suggest that a reasonable balance is found when everybody acting in good faith is protected from psychological bias.

For the current calculation, we believe the chosen blinding procedure to be successful in that regard. However, it came at the cost of a $\pm 0.0025$ level uncertainty, limiting the optimization of our preferred procedure.  This uncertainty is introduced by the $a^4$ terms in Eq.~\eqref{eqn:blind} that are not always eliminated by the continuum extrapolation.  The analysis groups, however, had to make decisions and freeze their analyses based on the blinded data set.  In Fig.~\ref{fig:effectblinding}, we highlight this effect by contrasting the relative unblinding as performed on the blinded data sets compared to the case, where we re-run the unmodified analyses on the unblinded data sets. 

In future studies, we will have to reconsider our exact approach since adding even higher-order terms (such as $a^6$) with sufficiently small coefficients to account for additional finer data sets would have a diminishing effect.  We may therefore decide to use only lattice-spacing-independent blinding factors in the future.

\section{Conclusions and Outlook}\label{sec:conclude}
In this work we compute the standard Euclidean window of the hadronic vacuum polarization.  We employ a blinded setup to avoid a possible bias towards reproducing previously published results.  We focus on the dominant quark-connected light-quark isospin-symmetric contribution and significantly improve its continuum extrapolation and address additional sub-leading systematic effects from sea-charm quarks and residual chiral-symmetry breaking from first principles.  
Our result for the total intermediate window $a_\mu^{\rm W}$ is in $3.8\sigma$ tension with the recently published dispersive result of Ref.~\cite{Colangelo:2022vok} and in agreement with other lattice results \cite{Borsanyi:2020mff,Ce:2022kxy,Alexandrou:2022amy}.  For the isospin-symmetric connected up and down quark contribution $a_\mu^{\rm W,iso,conn,ud}$ more lattice results are available \cite{Borsanyi:2020mff,Lehner:2020crt,Aubin:2022hgm,Wang:2022lkq,Ce:2022kxy,Alexandrou:2022amy} that are all in agreement with the result presented in this work.  

The tension for the intermediate window between lattice QCD and the dispersive result needs to be addressed in future work and a systematic study of additional windows may provide further insights.  As it stands, this tension may be interpreted as a yet to be understood new physics contribution to hadronic $e^+ e^-$ decays.  In the context of the 4.2$\sigma$ tension for $a_\mu$ \cite{Muong-2:2021ojo},
\begin{align}
  a_\mu({\rm EXP}) - a_\mu({\rm SM}) = 25.1(5.9) \times 10^{-10} \,,
\end{align}
we note that the difference of the dispersive and lattice results for $a_\mu^{\rm W}({\rm SM})$ is only $6 \times 10^{-10}$.

In addition, we provide a result for the short-distance window for which our result is compatible with the recently published result of the ETMC collaboration \cite{Alexandrou:2022amy}.  At short distances, we contrast lattice QCD and perturbative QCD and find agreement up to $0.4$ fm at the level of $1.5 \times 10^{-10}$.  We also provide results for a hybrid method in which we use perturbative QCD below $0.1$ fm and lattice QCD at longer distances.
The effective mass-independence of the vector correlators at short distances finally motivates us to define a mass-splitting procedure to further improve the continuum extrapolation of the HVP.

We are currently generating additional ensembles with lattice spacings at $a^{-1} = 3.5$ GeV and $4.7$ GeV that will support a five-lattice spacing continuum extrapolation using the mass-splitting method.

Finally, we are also preparing an update for the long-distance window using the improved bounding method \cite{Bruno:2019nzm} and an update of our QED and strong-isospin-breaking corrections re-using data from our hadronic light-by-light program \cite{Blum:2015gfa,Blum:2016lnc,Blum:2017cer,Blum:2019ugy}.
Upon completion of our HVP program, we expect to be able to match the FNAL E989 target precision.
 
\section{Acknowledgments}
We thank our colleagues of the RBC and UKQCD collaborations for many valuable discussions and joint efforts over the years.
The authors gratefully acknowledge the Gauss Centre for Supercomputing e.V. (www.gauss-centre.eu) for funding this project by providing computing time on the GCS Supercomputer JUWELS at Jülich Supercomputing Centre (JSC).
An award of computer time was provided by the ASCR Leadership Computing Challenge (ALCC) and Innovative and Novel Computational Impact on Theory and Experiment (INCITE) programs. This research used resources of the Argonne Leadership Computing Facility, which is a DOE Office of Science User Facility supported under contract DE-AC02-06CH11357. This research also used resources of the Oak Ridge Leadership Computing Facility, which is a DOE Office of Science User Facility supported under Contract DE-AC05-00OR22725.
This research used resources of the National Energy Research Scientific Computing Center (NERSC), a U.S. Department of Energy Office of Science User Facility located at Lawrence Berkeley National Laboratory, operated under Contract No.~DE-AC02-05CH11231 using NERSC award NESAP m1759 for 2020. This work used the DiRAC Blue Gene Q Shared Petaflop system at the University of Edinburgh, operated by the Edinburgh Parallel Computing Centre on behalf of the STFC DiRAC HPC Facility (www.dirac.ac.uk). This equipment was funded by BIS National E-infrastructure capital grant ST/K000411/1, STFC capital grant ST/H008845/1, and STFC DiRAC Operations grants ST/K005804/1 and ST/K005790/1. DiRAC is part of the National E-Infrastructure. 
We gratefully acknowledge disk and tape storage  provided by USQCD and by the University of Regensburg with support from the DFG.
The lattice data analyzed in this project was generated using GPT \cite{GPT}, Grid \cite{GRID}, and CPS \cite{CPS} and analyzed, in part, using pyobs \cite{PYOBS}.
TB is supported by the US DOE under grant DE-SC0010339.
PB, TI, CJ, and CL were supported in part by US DOE Contract DESC0012704(BNL), and PB, TI, and CJ were supported in part by the Scientific Discovery through Advanced Computing (SciDAC) program LAB 22-2580.
The research of MB is funded through the MUR program
for young researchers ``Rita Levi Montalcini''.
This project has received funding from Marie
Sk{\l}odowska-Curie grant 894103 (EU Horizon 2020).
VG and RH are supported by UK STFC Grant No. ST/P000630/1.
NM is supported by the Special Postdoctoral Researchers Program of RIKEN. 
TI is also supported by the Department of Energy, Laboratory Directed Research and Development (LDRD No. 23-051) of BNL and RIKEN BNL Research Center.
LJ acknowledges the support of DOE Office of Science Early Career Award DE-SC0021147 and DOE grant DE-SC0010339.
RM is supported in part by the US DOE under grant DE-SC0011941.
The work of ASM was supported by the Department of Energy, Office of Nuclear Physics, under Contract No. DE-SC00046548.
\bibliography{references}

\end{document}